\documentclass[printer]{aa} 
\usepackage{graphicx}
\usepackage{epstopdf}
\usepackage{txfonts}
\usepackage{natbib}
\bibpunct{(}{)}{;}{a}{}{,} 

\def \aa {A$\&$A$\;$}

\def \degmark{^\circ}

\def \phcmsec{\hbox{photons cm$^{-2}$ s$^{-1}$}}

\def \enmev {$E > 100 \: MeV$}

\def \abe{ }
\def \abf{   }
\def \abg{ }
\def \abl{ }
\def \abh{ }
\def \mt{  }

\begin{document}
%

\title{Evaluating the Maximum Likelihood Method
for Detecting Short-Term Variability of AGILE $\gamma$-ray Sources}
   \subtitle{}

\author{A.~Bulgarelli\inst{1}, A.W.~Chen\inst{3},
  M.~Tavani\inst{2,4},
  F.~Gianotti\inst{1}, M.~Trifoglio\inst{1}, T.~Contessi\inst{3}
}

\institute{$^1$INAF/IASF--Bologna, Via Gobetti 101, I-40129 Bologna, Italy \\
$^2$INAF/IASF--Roma, Via del Fosso del Cavaliere 100, I-00133 Roma, Italy \\
$^3$INAF/IASF--Milano, Via E.~Bassini 15, I-20133 Milano, Italy \\
$^4$Dip. di Fisica, Univ. ``Tor Vergata'', Via della Ricerca Scientifica 1, I-00133 Roma, Italy \\
}

   \date{received; accepted}

\authorrunning {A.Bulgarelli}
\titlerunning { Evaluating Maximum Likelihood 
for AGILE Short-Term Variability }
\offprints{A. Bulgarelli, \email{bulgarelli@iasfbo.inaf.it} }

\abstract{\mt The AGILE space mission (whose instrument is
sensitive in the energy ranges 18-60 keV, and 30 MeV - 50 GeV) has
been operating since 2007. Assessing the statistical significance
of time variability of $\gamma$-ray sources above 100 MeV is a
primary task of the AGILE data analysis. In particular, it is
important to check the instrument sensitivity in terms of Poisson modeling of the data background, and to
determine the post-trial confidence of detections. 
}
{\mt The goals of this work are: (i) evaluating the distributions of the likelihood ratio test for
"empty" fields, and for regions of the Galactic plane; 
(ii) \abl{calculating the probability of false detection over multiple time intervals}.}{\mt In this paper we describe  in detail the techniques
used to search for short-term variability in the AGILE $\gamma$-ray
source database. We describe  the binned maximum likelihood method
used for the analysis of AGILE data, and the numerical
simulations that support the characterization of the statistical
analysis. We apply our method to both Galactic and extra-galactic
transients, and provide a few examples. }{ After having checked
the reliability of the statistical description tested with the
real AGILE data, we obtain the distribution of $p$-values for blind
and specific source searches. We apply our results to the
determination of  the post-trial statistical significance of
detections of  transient $\gamma$-ray sources in terms of  pre-trial
values. }{The results of our analysis allow a precise
determination of the post-trial significance of $\gamma$-ray sources
detected by AGILE. }


   \maketitle
%

\section{Introduction}
\label{sec:intro}

{\mt  The current generation of $\gamma$-ray space missions, AGILE
and Fermi, are sensitive in the energy range above 100 MeV up to
tens of hundreds of GeV. These missions are
providing a great wealth of data on a variety of $\gamma$-ray
sources, both in our Galaxy and at extragalactic distances.
Compared with the previous generation of $\gamma$-ray instruments
(e.g., EGRET on board of the Compton Gamma-Ray Observatory), 
the AGILE and Fermi-LAT $\gamma$-ray imagers are
based on silicon detectors with optimal spatial resolution and
much improved background rejection. These characteristics allow
reaching a few arcminute positioning for intense sources and very
large fields of view (FOVs). Both AGILE and Fermi-LAT reach  FOVs
more than 100 degrees across, i.e., 2.5 sr, and this fact is of
great relevance for the statistical analysis of the $\gamma$-ray
sources. Based on the current detector performances, it is crucial
to perform a statistical analysis specialized to the specific
modes of operation of the two $\gamma$-ray missions. In particular,
the distinction between non-significant and significant
steady/transient sources must be supported by a specific
treatment of the satellite $\gamma$-ray data. The relatively large
effective areas at $\gamma$-ray energies, and the very large FOVs,
produce large exposures (measured as the product of effective area
times the on-source duration). The statistical significance of
$\gamma$-ray source detections must address the pre- vs. post-trial
significance. As we will see, the characteristics
of the pointing and of the specific search (whether a "blind"
search or a search for a specific source) have a direct influence
on the proper statistical treatment. In this paper, we address the
issue of the statistical determination of $\gamma$-ray sources for
the AGILE mission.

}

\abh{The AGILE-GRID instrument for observations in the $\gamma$-ray has an energy
range of 30~MeV - 50~GeV \citep{Tavani_2009a}}. AGILE  data are down-linked approximately
every 100 minutes and sent to the AGILE Data
Center (ADC), which is part of the ASI Science Data Center (ASDC) for data reduction, scientific processing, and archiving. 
ASDC forwards the AGILE data to the AGILE Team local sites where a Quick Look analysis is performed.

Because of the low detection rate of events and the extent of the  AGILE GRID point-spread function (PSF),
statistical techniques like the maximum likelihood method (or estimator) are required to analyze the AGILE point sources.
Similar analysis techniques were used for EGRET data \citep{Mattox_1996a}.

For steady sources and sources with high $\gamma$-ray flux, the significance of a detection increases as a function of the observation duration. This is not necessarily the case for variable sources with low flux:   a short integration time (the integration period should be of the same time scale as the duration of the astronomical phenomena under study) may yield a low significance level because the observation is \textit{photon limited}; for a longer integration time the source may disappear entirely because the observation is \textit{background limited}.

The main purpose of this work is the evaluation of the likelihood ratio test in the context of short time scale (1-2 days) flaring $\gamma$-ray sources. The likelihood ratio test is used to compare two ensembles of hypotheses, one in which a $\gamma$-ray source is present, and another (the null hypothesis) in which it is absent. The determination of the likelihood ratio distribution in the case of the null hypothesis is used to evaluate the occurrence of false positive detections.

The search for $\gamma$-ray transients (Galactic and
extra-Galactic) detectable on timescales of 1-2 days is one of the
 major activities performed by the AGILE Collaboration\footnote{ Two
independent $\gamma$-ray transient search systems have been
developed. One system operates  at
INAF/IASF-Bologna \citep{Bulgarelli_2009}; it is able to process
the data  {\mt within} 1.5 hours (from the last photon acquired in
orbit to the alert generation), and to generate alerts via e-mail
and SMS to the mobile phones of the AGILE Collaboration. A second system
operates at the ADC (AGILE Data Center) in Frascati; it can
react within an average time of 3 hours, and generates alerts via
e-mail but with more accurate data processing.}.

The {\mt main goals}
of this work are: (i) to evaluate the  distribution of the
likelihood ratio \abg{test statistic} ($T_s$) for empty fields both outside of and
within the Galactic plane; (ii) to evaluate the $T_s$ distribution
in the case of one flaring source with different combinations of
parameters;  
and
(iii) to calculate the  confidence levels corresponding to the same
\abg{false detection probability} in multiple non-overlapping time periods.

We performed \abh{Monte Carlo} simulations to determine the $T_s$
distributions both in the presence and absence of a flaring
source, which may differ from the theoretically predicted
distribution for a number of reasons which we
{\mt discuss} below. We compare these results with the AGILE data
where feasible, and show {\it how the formulation of the
hypotheses influences the $T_s$ distribution}.

In Section \ref{sec:ana} we present the AGILE{\mt -GRID} maximum
likelihood analysis method. In  Section
\ref{sec:montecarlo} we report a general description of the
{\mt performed} \abh{Monte Carlo} simulations. In Section \ref{sec:montecarlo1} we characterize  the $T_s$ distribution of a simulated extra-Galactic empty field and compare it with a real observation case. Section
\ref{sec:montecarlo2}  characterizes the $T_s$ distribution  for two simulated Galactic fields, a simple and a complex case; in the latter we analyze the effect \abl{that} uncertainties in the analysis parameters produce on the $T_s$ distribution.
In Section \ref{sec:pretrials} we describe the pre- and post-
trial probability, and in Section \ref{sec:multiple} we consider
multiple detections from the same sky position.


\section{Analysis method}
\label{sec:ana}

The \textbf{likelihood ratio test} is used to compare two
ensembles of models, one of which is a subset of the other, each
of which can be characterized by a set of parameters. In the most
common case, one of the ensembles of models is the null
hypothesis, while the other, of which the null hypothesis is a
subset, is the alternative hypothesis, corresponding to, for
example, the hypothesis of the existence of a source. In each
case, the values of the set of parameters are found \abe{by means
of a maximum likelihood method or estimator} which maximizes the
\textbf{likelihood} of producing the
data given the ensemble of models. The application of likelihood to photon-counting
experiments is described in \citep{Cash_1}. Details of how the likelihood
is calculated {\mt in the context of $\gamma$-ray data analysis} can
be found in \citep{Mattox_1996a}.
The likelihood ratio is then simply the ratio of these two maximum
likelihoods, and the test statistic $T_s$ is defined as
\begin{equation}
T_s = -2 ln \frac{L_0}{L_1}
\label{eqn_TS}
\end{equation}
where $L_0$ and $L_1$ are the maximum value of the likelihood
function for the null hypothesis and for the alternative
hypothesis, {\mt respectively}.

In the AGILE-GRID case, the data are binned into FITS counts  maps, while each model is
a linear combination of isotropic and Galactic diffuse components of the $\gamma$-ray emission
 and point sources.
 {\mt $\gamma$-ray exposure maps, and galactic diffuse emission maps
 are used to calculate the models}. Among the parameters which may be varied to
 find the maximum likelihood are the coefficients of the diffuse and point source components.

\abg{In general, } to describe a single point source, four parameters are used: \abg{the predicted source counts} $s_c$,
the spectral index $s_{si}$ \abg{(the spectrum of each source is assumed to be a power-law)}, and two parameters corresponding to
the position of the source ($s_l, s_b$, in Galactic coordinates). 
It is
possible to keep each parameter free or fixed; a free parameter is
allowed to vary to find the maximum likelihood. Possible
combinations include: only allowing the flux to vary, allowing
both the position and flux to vary, allowing both the spectrum and
flux to vary, or leaving all four parameters free. \abh{Throughout
this paper we keep the spectral index fixed to 2.1; for this kind of sources the AGILE-GRID instrument has a  point spread functions (PSF) at 30$\degmark$  off-axis   for E$>$100 MeV of 2.1$\degmark$, for E$>$400 MeV of 1.1$\degmark$, for E$>$1 GeV of 0.8$\degmark$.}

The two parameters that describe the Galactic (diffuse) and
isotropic $\gamma$-ray emission are: (i)  $g_{gal}$, the
coefficient of the Galactic diffuse emission model, and (ii)
$g_{iso}$, the isotropic diffuse intensity. A value of $g_{gal}
\le 1$ is expected if the galactic diffuse emission model is
correct. Values of $g_{iso} $ between 1 and 15 $\times 10^{-5}
$cm$^{-2}$ s$^{-1}$ sr$^{-1}$ are expected, depending on pointing
strategy and on board and background filter rejection. These two
parameters can be left free or fixed independently. For short
timescale variability (less then 3-4 days) of $\gamma$-ray
sources, usually we first estimate these parameters with a longer
timescale integration, and then fix them for the short timescale
analysis, assuming that these components do not vary significantly
on the shorter timescales. We note that for {\mt the AGILE data
used in this analysis (based on the standard filter FM3.119)}, the isotropic component is dominated by instrumental
charged particle background rather than {\mt by } the
extragalactic diffuse emission, in contrast to data from EGRET and
Fermi-LAT.

The values of the parameters which maximize the likelihood are
those which describe the model in the ensemble most likely to
reproduce the data.

The null hypothesis corresponds to the
absence of the point source,
 while the alternative hypothesis corresponds to its presence. Clearly,
 the null hypothesis is a
subset of the alternative hypothesis, corresponding to a source
with zero flux. The Galactic diffuse and isotropic coefficients,
as well as the parameters of other known point sources in the
field of view, must be kept either fixed or free in the same
manner when evaluating both the null and alternative hypothesis.

In order to limit the effect of systematic errors far from the position
of the hypothesized source, the data bins evaluated
(and their predicted values according to the models) are limited to
an \textbf{analysis region} of radius $5\degmark$ or $10\degmark$ centered
around the source position.

From Wilks's Theorem \citep{Wilks_1938} the $T_s$ distribution $
\varphi $ \abg{is expected to asymptotically follow}
$\chi^{2}_{n-m}$ in the null hypothesis, where $n-m$ is the number
of additional parameters that are optimized in the alternative
hypothesis. In the most simple case, $n-m=1$ (e.g., in the case of
the determination of the flux of a single source). This means that
from Wilks's theorem, $T_s$ is expected to be asymptotically
distributed as $\chi^{2}_{1}$ in the null hypothesis. The expected
departure of the distribution from $\chi^{2}_{1}$ is of order
$(N)^{-1/2}$ where N is the number of \textbf{samples}. In our
context, the number of samples \abl{is} the number of photons which
carry information about all the parameters; these are all the
photons in the analysis region. This is true regardless of the
number that \abl{is} eventually estimated to come from the point
source.

When there are multiple sources whose fluxes are allowed to vary,
the following procedure, \abl{divided into   two loops}, is used to find the maximum $T_s$. \abl{In the first loop, first} 
the sources are sorted according to hypothesized flux. One by one,
the sources are added to the model, from highest to lowest flux.
If the source flux is allowed to vary, then the maximum likelihood
is found both in the presence and absence of the source. If the
position is allowed to vary, \abl{the first fit is done at a fixed position and} the resulting $T_s$ is compared to a
location confidence level threshold ($t_{lcl}$). If $T_{s} >
t_{lcl}$, then $T_{s}$ is again maximized with variable flux and
position. \abl{$T_{ss}$ is a threshold for
promoting the source to the second loop of the algorithm:} if the final $T_{s}$ is greater than 
$T_{ss}$, the source is considered significant and added to both
the null and alternative hypotheses for the other sources. If not,
it is considered undetected, and is set to zero flux for all
subsequent analyses. \abg{With the sources over the $T_{ss}$ threshold (with or without a location confidence level)} \abl{the} second loop is similar to the
first loop, except that all of the sources marked significant in
the first loop are contained in the models from the beginning. The
sources are again evaluated one at a time from highest to lowest
flux. The $T_s$ of each source is again maximized, and set to its
final value.
The values of the parameters $t_{lcl}$ and $T_{ss}$ affect the behavior of the procedure.
$t_{lcl}=5.99147$ corresponds to a $95\%$ confidence level for two degrees of freedom.


The maximum likelihood estimator developed for AGILE constrains the flux of a source,
and therefore the source counts $s_c$, to be greater than or equal to zero.
Because the ensemble of models considered is half of the theoretically possible number,
the shape of the $T_s$ distribution differs from that of Wilks's theorem
by being asymptotically distributed as $0.5 \times \chi^{2}_{1}$ instead of $\chi^{2}_{1}$ \citep{Mattox_1996a}.

In order to compare the  data distribution of $T_s$
produced by the AGILE analysis procedure {\mt with} that predicted
by Wilks's theorem, we performed a series of \abh{Monte Carlo}
simulations of AGILE data. Each simulation of an analysis region
and its subsequent maximum likelihood analysis constitutes a
single {\mt trial}.
The probability that the result of a trial in an empty field has
$T_s \geq h$ (that is the complement of the cumulative
distribution function) is

\begin{equation}
P(T_s \geq h) = \int_{h}^{+\infty} \varphi (x) dx
\label{eq_A}
\end{equation}

This is also called the $p$-value $p=P(T_s \geq h)$. This is the
pre-set (pre-trial) type-1 error (a false positive, rejecting the
null hypothesis when in fact it is true). Given a statistical
distribution, a "$p$-value" assigned to a given value of a random
variable is defined as the probability of obtaining that value or
larger when the null hypothesis is true. This value may be
interpreted as an "occurrence-rate", that is, how many trials
occur on average before obtaining a false detection at a level
equal or greater to $h$.

\subsection{Hypothesis formulation}

In the context of the $\gamma$-ray transient analysis, the null
hypothesis is defined as an analysis region containing only steady
and known sources with no flaring sources present. We can
translate this into the ensemble of models by keeping the flux of
the flaring source fixed to zero, and the \abl{fluxes} of steady sources
fixed to their known fluxes. In the alternative hypothesis that a
flaring source is present, the flux  (and position if specified)
of this source is allowed to be free and the \abh{fluxes} of steady
sources are fixed to their known fluxes. In this work, we restrict
our analysis to hypotheses of single flaring sources, neglecting
alternative hypotheses of 2 or more flaring sources in the same
analysis region.

Additional knowledge of the source, e.g. from other wavelengths,
can add useful additional constraints about the position of the
source in the hypothesis formulation. In Section
\ref{sec:montecarlo2} we show that this additional knowledge can
change the $T_s$ distribution, and thereby reduce the occurrence
rate of false detection.

For the analysis of a flaring source, we consider two possible scenarios:
\begin{enumerate}
\item The flaring source is unknown: in this case, the position
and flux parameters are allowed to be free and optimized with
respect to the input data: the starting $(l,b)$ position is
usually the counts peak found in the smoothed map. If the $T_s$ is
higher than a well defined threshold, the alternative (flaring
source) hypothesis is accepted, and a counterpart search may be
performed. \item The source is known and the alternative
hypothesis is that this source is in flaring state. This scenario
can be further subdivided into two cases:
    \begin{enumerate}
    \item  The mean flux of the source is below the background level of the sky region,
    producing only an upper limit over long integrations.
    \item  The mean flux of the source \abl{is} above the background level of the sky region
    and is detected over long integrations.
    \end{enumerate}
In both sub-cases, two types of analysis are possible: (i) the
flux parameter is allowed to be free and the position kept fixed,
(ii) the flux and position parameters are allowed to be free. If
the position is allowed to be free, the starting position is
usually the position of the source in steady state. Keeping the
position fixed implies that the alternative hypothesis being
tested is that the flare comes from the known source (e.g. from
the behaviour in other wavelengths or because other flares have
been detected in the past), whereas allowing the position to vary
allows the alternative hypothesis to include any flaring source
within the analysis region. We may then, at the same time as we
calculate the significance of the detection, verify whether the
confidence contour of the source position is compatible with the
source hypothesized to be responsible for the flare. If the
position of the known source is outside the confidence region then
the alternative hypothesis can be rejected in the sense that the
known source is not responsible for the flare.
\end{enumerate}

\section{Monte Carlo simulations}
\label{sec:montecarlo}

Monte Carlo simulations of AGILE {\mt $\gamma$-ray } data were used
to characterize the maximum likelihood analysis procedure.
Simulated data were generated using a model of the background
(Galactic diffuse radiation model and isotropic background) and
the AGILE-GRID instrument response functions (version I0023 of the
calibration matrices for effective area, energy dispersion and
point spread function). \abg{The energy range used is 100 MeV - 50 GeV.} The simulated observations were generated
adding Poisson-distributed deviates to each pixel. Each bin of the
generated maps (counts, exposure and  Galactic emission maps) has
been analyzed exactly as flight data  as described in Section
\ref{sec:ana}. The bin size chosen for the simulations is
$0.25\degmark$, the same size used in the {\mt AGILE-GRID} daily
monitoring.

The exposure level chosen for the simulations was a level equivalent
to a mean value of 1-day pointing/2-day spinning AGILE observation mode.

In Table \ref{table_ef33} we report the parameters and the number of trials for the performed simulations.

\begin {table*}[!htb]
\caption {\em{Parameters and number of trials for the performed extra-Galactic and Galactic field simulations.}}
\label{table_ef33}
\renewcommand{\arraystretch}{1.2} 
\begin{tabular}{|c|c|c|c|c|c|}
\hline & & &  Simulated & Analyzed & \\
simulation & source position & $t_{lcl}$  & ($g_{gal}$ , $g_{iso}$) & $g_{gal}$ and $g_{iso}$ & Trials $\times 10^6$ \\
\hline Extra-Gal.  empty field & fixed & na & $(1.0, 6.0)$ & free & 31.0\\
\hline Extra-Gal.  empty field & fixed & na & $(1.0, 12.0)$ & free &  35.0 \\
\hline Gal.  field without sources & fixed & na & $(1.0, 6.0)$ &  free &   7.0 \\
\hline Gal.  field without sources & free & 5.99147  & $(1.0, 6.0)$ &  fixed &   4.0 \\
\hline Gal.  field with steady sources & fixed & na  & $(0.63, 7.70)$  & fixed &  9.0 \\
\hline Gal.  field with steady sources & free &  5.99147 & $(0.63, 7.70)$ & fixed & 17.0\\
\hline Gal.  field with steady sources & free &  2.29575 & $(0.63, 7.70)$ & fixed &  2.2\\
\hline
\end{tabular}
\end{table*}

\section{Extra-Galactic empty field }

\label{sec:montecarlo1}

\subsection{Monte Carlo simulation }

\begin{figure}[!htb]
\centering
\includegraphics[width=9 cm]{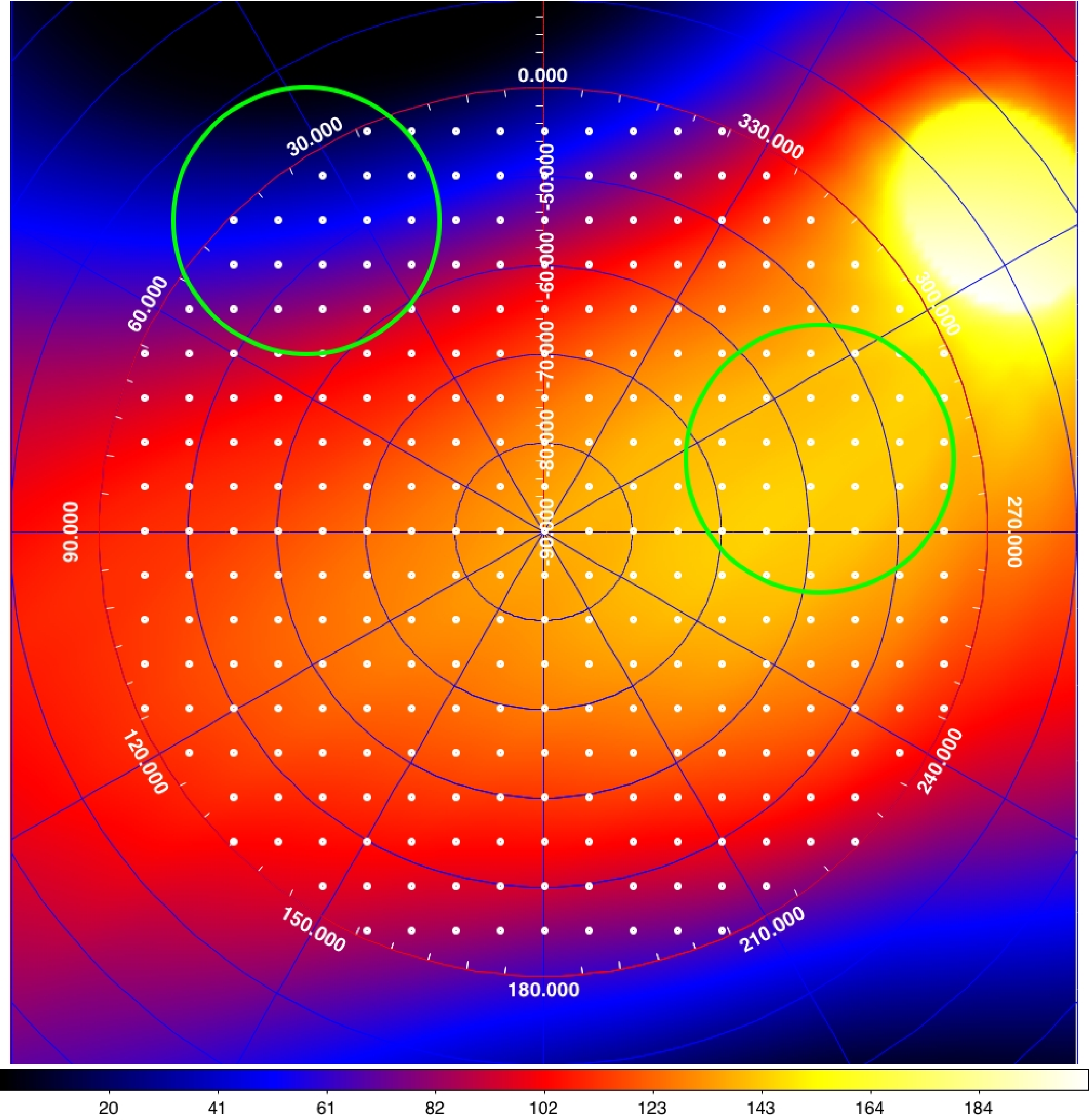}
\centering \caption { {\it Extra-Galactic ($b=90\degmark$)
exposure map used for the simulations {\mt (in units of $cm^2 s
sr$) in Galactic coordinates. Notice the larger exposure near the celestial pole. \abg{The white circles are the positions of the trials, the green circles indicate the high and low chosen exposure regions.}}} }
\label{fig_EXPEXGAL}
\end{figure}

We simulated an extra-Galactic empty field without flaring or
steady sources. The simulation was performed with an AGILE field
of view of $60\degmark$. Figure \ref{fig_EXPEXGAL} shows the
exposure map, and the  bins used in this simulation to perform the
trials. This is a typical 2-day exposure map in spinning mode, but
in this context the key point is the level of exposure, and not
its shape.  We analyzed positions within  $50\degmark$ of the
center of the map to exclude low values of the exposure, which we
also do in every day sky monitoring.

\begin{figure}[!htb]
\centering
\includegraphics[width=9 cm]{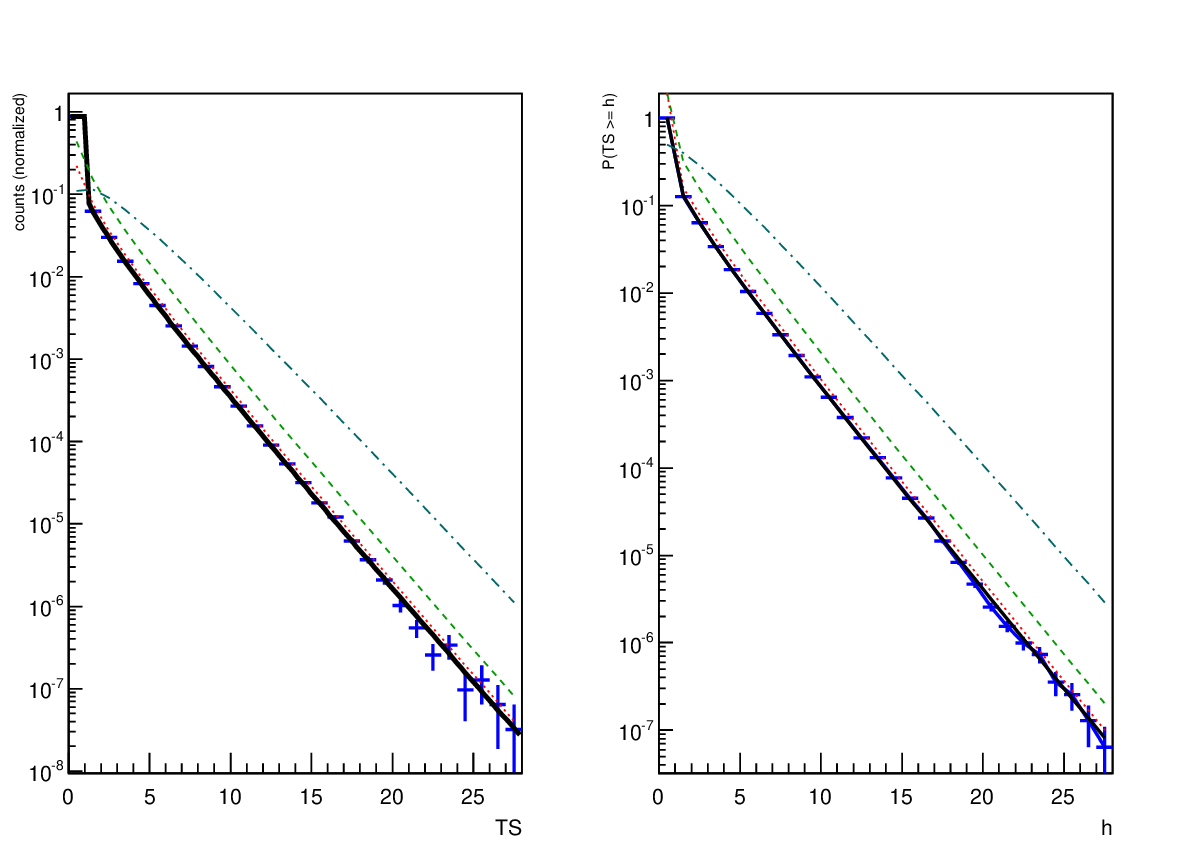}
\centering \caption { {\itshape $T_s$ distribution (left side) and  $p$-distribution (right side) of a simulated empty field, with $g_{gal}=1$ and $g_{iso}=6$ (with these parameters left free during the analysis), flux free, position fixed. The blue \abl{crosses are} the calculated distribution, the black line is the best fit according to Equation \ref{eqn_fit1}, the red dotted line is the $\frac{1}{2} \chi^2_1$ theoretical distribution, the green dashed line is the $\chi^2_1$  theoretical distribution, the Cyan dotted-dashed line is the $\frac{1}{2} \chi^2_3$ distribution.} }
\label{fig_F1}
\end{figure}

We performed a maximum likelihood analysis at positions
corresponding to every fifth \abg{degree} on the map. The spacing was
chosen to ensure that the analyses would be independent from one
another. The position \abg{is} kept fixed and the
flux allowed to vary, implying one additional parameter in the
alternative hypothesis.
We repeated the counts map simulations and analysis for different
values of the coefficient of the isotropic diffuse component
($g_{iso}$=6 and 12) consistent with  values found in real
AGILE observations.
Figure \ref{fig_F1} shows the resulting $T_s$ distribution
(left panel)
 and the related $p$-value distribution (right panel).
 We fit this $T_s$ distribution with the following function:

\begin{equation}
$$
\mbox{$\kappa'(T_s)$} = \left\{ \begin{array}{ll}
\mbox{ $\delta  $ } &\mbox{ if $T_s<1$} \\
 \mbox{ $\eta \chi^2_{N} (T_s)$ } &\mbox{ otherwise } \\
       \end{array} \right.
$$
\label{eqn_fit1}
\end{equation}

\noindent In Table \ref{table_ef2} we \abh{report} the results of the
fit for different  background $g_{iso}$ levels. The   function $\delta$
in the first bin takes into account the constraint on the source
counts ($s_c \ge 0$) . In this simple case, the simulated
distribution is \abl{close} to the expected $1/2 \chi^2_1$ (see the
$\eta$ parameter in Table \ref{table_ef2}).

\noindent \abg{In Table \ref{table_ef3} we report the results of the
fit for different levels of exposure.  Figure \ref{fig_EXPEXGAL} shows the 
regions with different exposure levels.}

\begin {table}[!htb]
\caption {\em{Best fit parameters in the case of an empty field
for simulated sky maps with bin size of $0\degmark.25$ for
different values of $g_{iso}$, the isotropic emission coefficient.
The fitting function is \abf{reported in Equation \ref{eqn_fit1}}.}}
\label{table_ef2}
\renewcommand{\arraystretch}{1.2} 
\begin{tabular}{|c|c|c|c|c|}
\hline $g_{iso}$     & $\delta$ & $\eta$  \\
\hline 6 &  0.8742  $\pm  1.9 \cdot 10^{-4}$ & 0.4082  $\pm  2.3 \cdot 10^{-4}$  \\
\hline 12  & 0.8681 $\pm  1.5 \cdot 10^{-4}$ & 0.4280  $\pm  2.0 \cdot 10^{-4}$  \\
\hline
\end{tabular}
\end{table}

\begin {table}[!htb]
\caption {\em{Best fit parameters in the case of an empty field
for simulated sky maps with bin size of $0\degmark.25$, $g_{iso}$=6, for
two levels of exposure.
The fitting function is \abf{reported in Equation \ref{eqn_fit1}}.}}
\label{table_ef3}
\renewcommand{\arraystretch}{1.2} 
\begin{tabular}{|c|c|c|c|c|}
\hline $exposure$     & $\delta$ & $\eta$  \\
\hline low &  0.8792  $\pm  5.6 \cdot 10^{-4}$ & 0.3920  $\pm  6.7 \cdot 10^{-4}$  \\
\hline high  & 0.8715 $\pm  5.3 \cdot 10^{-4}$ & 0.4172  $\pm  6.6 \cdot 10^{-4}$  \\
\hline
\end{tabular}
\end{table}

\subsection{Real   observation }
\label{sec:real1}

\begin{figure}[!htb]
\centering
\includegraphics[width=9 cm]{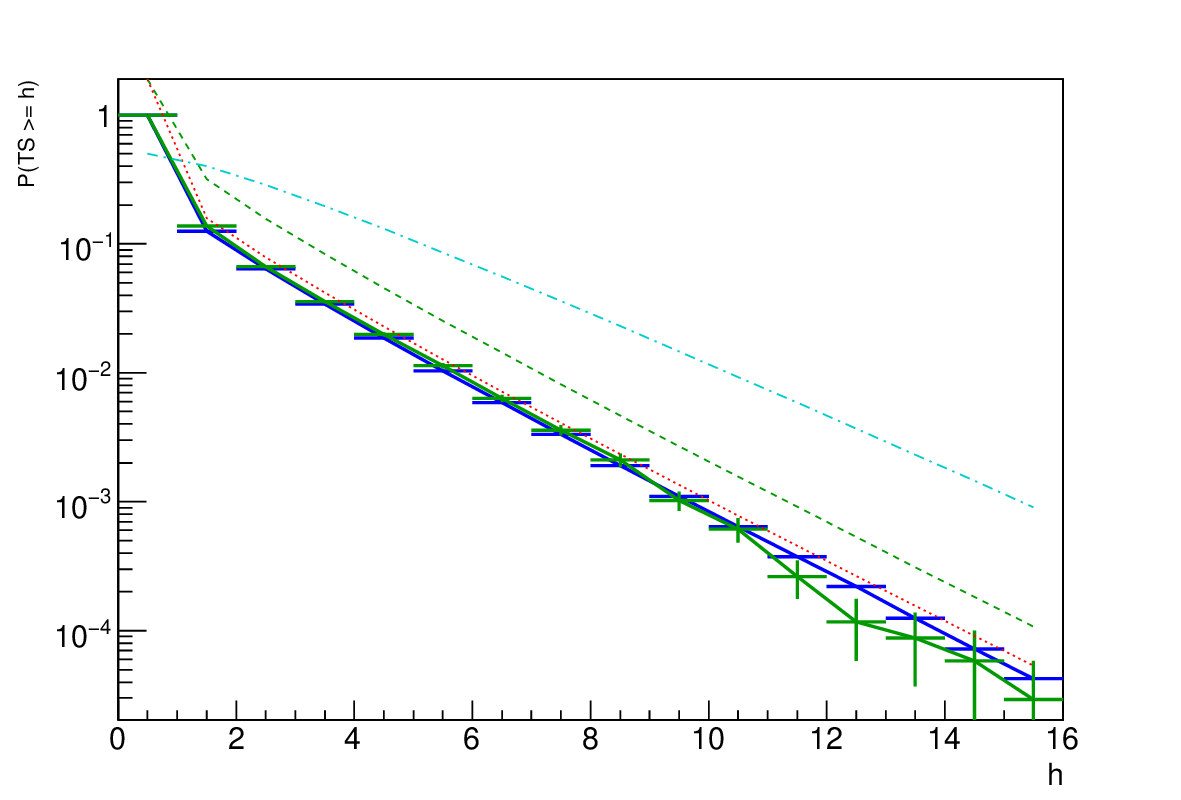}
\centering \caption { {\itshape  Comparison between {\mt
$p$-value} distributions for simulated (blue) and real (green)
empty extra-Galactic {\mt fields}. The red dotted line is the
$\frac{1}{2} \chi^2_1$ theoretical distribution, the green dashed
line is the $\chi^2_1$  theoretical distribution, the Cyan
dotted-dashed line is the $\frac{1}{2} \chi^2_3$ distribution.} }
\label{fig_FOB7410_4}
\end{figure}

We compared the simulated data shown in the last section to a real
AGILE observation. The observation block chosen is OB7410 (see ASI
Data Center web site, http://agile.asdc.asi.it/), in which AGILE
was pointed to the North Galactic Pole with good exposure. For
each day of the observation counts, exposure and gas maps were
generated and analyzed. As in the Monte Carlo simulations, a
maximum likelihood analysis was performed for a hypothetical
source position at every fifth \abl{degree} to ensure the independence of
each trial. 
%
The results are shown in Figure \ref{fig_FOB7410_4}.
Taking into account the limits of the statistics collected from this real observation,
real and simulated data are compatible at 1$\sigma$ error level.



\section{Monte Carlo simulations of Galactic fields }
\label{sec:montecarlo2}

We performed simulations of two regions {\mt of the Galactic
plane}: a Galactic region {\mt with a low density of potential
sources}, and a complex Galactic region (the Cygnus region). These
two regions represent two extremes {\mt for} the AGILE analysis.

\subsection{A simple Galactic region }

\label{sec:montecarlo3} We performed a simulation {\mt of a
relatively simple} Galactic region {\mt by assuming only the
Galactic diffuse and isotropic emissions  without steady or
flaring sources. This calculation was aimed at evaluating  the
photon density function and the $p$-value } distribution of the
AGILE likelihood maximum estimator in the presence of the Galactic
diffuse emission. We chose a region centered on (l,b)=(160,0)
(Galactic coordinates) with 1-day exposure level in pointing mode.
The parameters used in the simulation are $g_{gal}=1$ and
$g_{iso}=3$. During the analysis, \abf{the spectrum} of {\mt any}
hypothetical source is kept fixed.

\begin{figure}[!htb]
\centering
\includegraphics[width=9 cm]{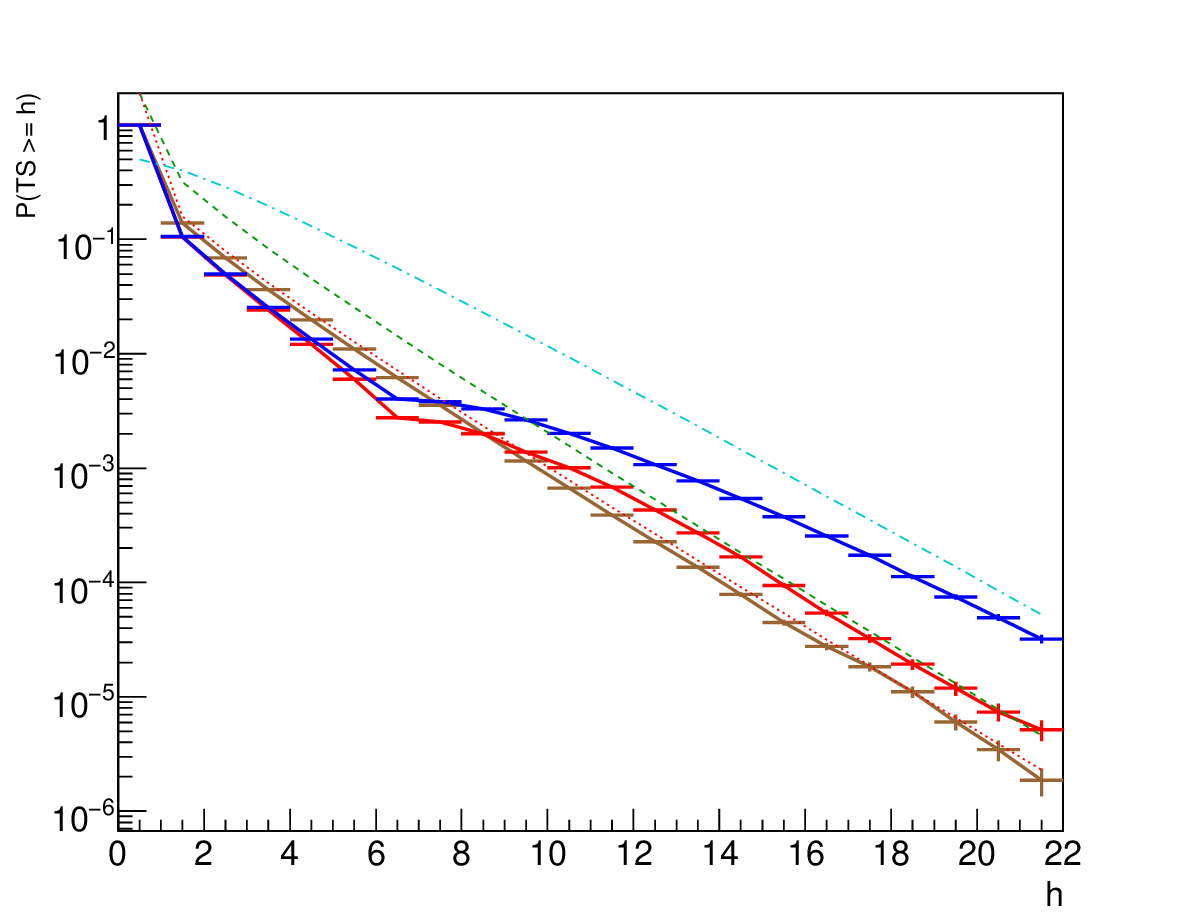}
\centering \caption { {\itshape The effect of the hypothesis
formulation (trial selection). The $p$-value distribution depends
on the constraint on the position of the source:  comparison
between different analysis methods (in particular, keeping fixed
and leaving free the position of the source) in the case of
absence of a source at the (l,b)=(160,0)$\degmark$ location.
Histograms are the  $p$-value distribution  for an empty Galactic
field when the null hypothesis   is true. Blue and red histograms
have the following parameters:  $T_{ss} =4$, $t_{lcl}=5.99147$, flux
and position of the source left free. The blue histogram contains
all trials regardless of the calculated position, while the red
histogram contains the trials that respect the ICL criterion. The
brown histogram has the position of the source kept fixed and flux
free. The red dotted line is the $\frac{1}{2}\chi^2_1$ theoretical
distribution, the green dashed line is the $\chi^2_1$  theoretical
distribution, the cyan dotted-dashed line is the $\frac{1}{2}
\chi^2_3$ distribution.} } \label{fig_OB2400_1BB}
\end{figure}

In order to analyze the flux of a source whose position is known,
{\mt we fix} the position of the source, and  allow the flux to
vary in the alternative hypothesis. The resulting $p$-value
distribution is shown as the brown histogram in Figure
\ref{fig_OB2400_1BB}. Using Equation \ref{eqn_fit1} we find the
best fit with
 $\delta= 0.8600 \pm 0.0003$
 and $\eta=0.4540 \pm 0.0005$,
for $N_1 = 1$. However, because of the presence of systematic
errors in the event reconstruction, there are cases in which the
source is detectable but at a position farther from the known
position than a purely statistical analysis would predict. In
order to handle these cases, we kept the position of the source
free and we have developed an analysis criterion which we call ICL
(Inside Confidence Contour Level): if the contour level is
present,  $T_s \geq t_{ICL}$ (we fix $t_{ICL} = 9$ for the AGILE
analysis), and the position of the source under investigation is
outside the contour \abl{level} found by the maximum likelihood
procedure, we reset $T_s$ to 0; \abg{a contour in
principle always exists (even if it is not necessarily closed or
connected), but  sometimes our software fails to find it.} \abg{The contour is always searched for at $t_{lcl}$.}

\abg{We use the ICL criterion only in the case of the presence of a known source. The $T_s$ distribution presented hereafter with position free and without ICL criterion are related to the analysis of an unknown source. The reason for throwing out the event is that the technique is used to weed out detections which we are not sure are
coincident with the source.}

\abl{We compare the results with the ICL criterion with the standard analysis (position left free)}. The
$p$-value distributions are reported in Figure
\ref{fig_OB2400_1BB} (the red histogram for the ICL criterion, the
blue histogram without the ICL criterion, \abl{both with $T_{ss} =4$, $t_{lcl}=5.99147$}). \abl{The blue and red histograms show a pronounced dip just above $T_{lcl}$ with
respect to the brown histogram because sources with $T_s  >  T_{lcl}$ 
may increase their $T_s$ during relocalization (shift to the right).} We characterize the
$T_s$ distribution produced by the addition of the ICL criterion
because it is used by the AGILE automated quick-look analysis when
searching for flares from known sources.

\begin{figure}[!htb]
\centering
\includegraphics[width=9 cm]{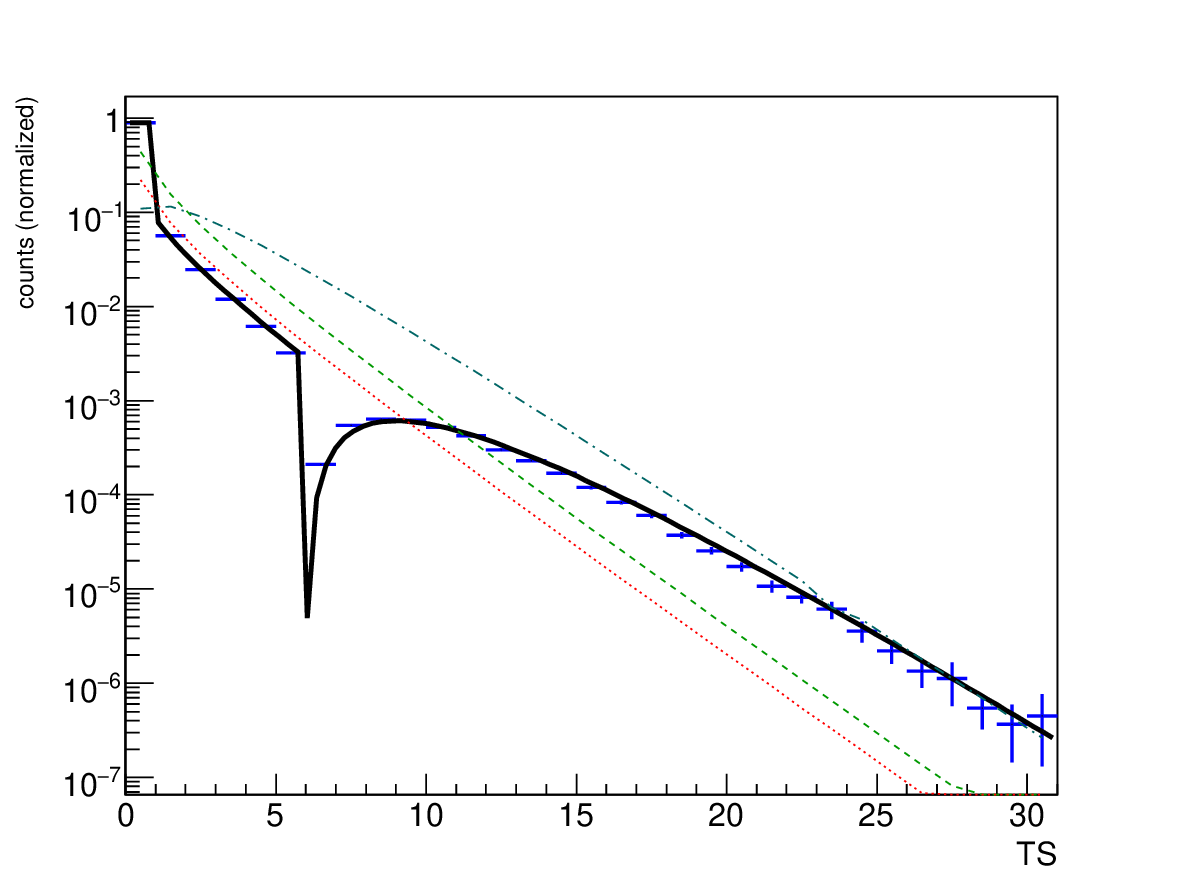}
\centering \caption { {\itshape The blue histogram is the $T_s$
distribution for the empty Galactic region when the null
hypothesis for a source at the position (l,b)=(160,0)$\degmark$ is
true. $T_{ss} =4$, $t_{lcl}=5.99147$, flux and position of the
source left free. Black line is the best fit function described in
Equation \ref{eqn-posfree}, $N_2=5$. The red dotted line is the
$\frac{1}{2}\chi^2_1$ theoretical distribution, the green dashed
line is the $\chi^2_1$  theoretical distribution, the cyan
dotted-dashed line is the $\frac{1}{2} \chi^2_3$ distribution.} }
\label{fig_OB2400_1A}
\end{figure}


Figure \ref{fig_OB2400_1A} reports the $T_s$ distribution when the
null hypothesis is true produced by maximum likelihood analysis
with the following parameters: $T_{ss} =4$, $t_{lcl}=5.99147$, flux
and position of a source at (l,b)=(160,0)$\degmark$ position left
free. The related $p$-value distribution has been already shown in
Figure \ref{fig_OB2400_1BB} (blue histogram). The ICL criterion is
not applied, i.e. no rejection is applied on the basis of the
compatibility of the location contour with the position of the
source. Fitting this {\mt distribution} with the following
function,
\begin{equation}
$$
\mbox{$\kappa''(T_s)$} = \left\{ \begin{array}{ll}
\mbox{ $\delta$ } &\mbox{ if $T_s<1$} \\
 \mbox{ $\eta_1 \chi^2_{N_1}(T_s)$ } &\mbox{ if $T_s\ge 1$ and $T_s\le t_{lcl}$ } \\
 \mbox{ $\eta_2 \chi^2_{N_2} (T_s - t_{lcl})$ } &\mbox{ otherwise } \\
       \end{array} \right.
$$
\label{eqn-posfree}
\end{equation}


\noindent we find  $N_1=1$ (if $T_s < T_{ss}$, no optimization of
the position takes place and therefore the only free parameter is
the flux of the source), $N_2=5$, $\delta =0.89 \pm 4.5\times 10^{-4}$,
$\eta_1 = 0.35 \pm 5.1\times 10^{-4}$ and $\eta_2 = 3.96 \times 10^{-3} \pm 3
\times 10^{-5}$. \abl{We use functions with N=5
(dof) solely as an analytical approximation to the functional form
produced by this process}. The translation $(T_s-t_{lcl})$ is due to the
switch between the fixed and free position regime \abl{(see Sect.
\ref{sec:ana})}.

Equation \ref{eqn-posfree} is appropriate for blind searches for
unknown sources. \abf{When searching for flares from a known
source (i.e. our hypothesis is {\mt that} there is a known source
at (l,b)=(160,0)$\degmark$ position)}, the appropriate alternative
hypothesis should exclude sources for which the known source
position lies outside the location contour. Therefore, applying
the ICL rejection criterion at the $95\%$ confidence level, the
resulting $p$-value distribution is reported in Figure
\ref{fig_OB2400_1BB} (red line) compared with the blue histogram
of the same Figure in which no selection criterion is applied. The
effect of appropriate hypothesis formulation is evident. The
application of {\mt the} ICL rejection reduces the number of
degrees of freedom. When we fit this distribution with Equation
\ref{eqn-posfree}  we find that the histogram has a distribution
between $N_2=3$ and $N_2=4$, due to the ICL selection criterion.


%

\begin{figure}[!htb]
\centering
\includegraphics[width=9 cm]{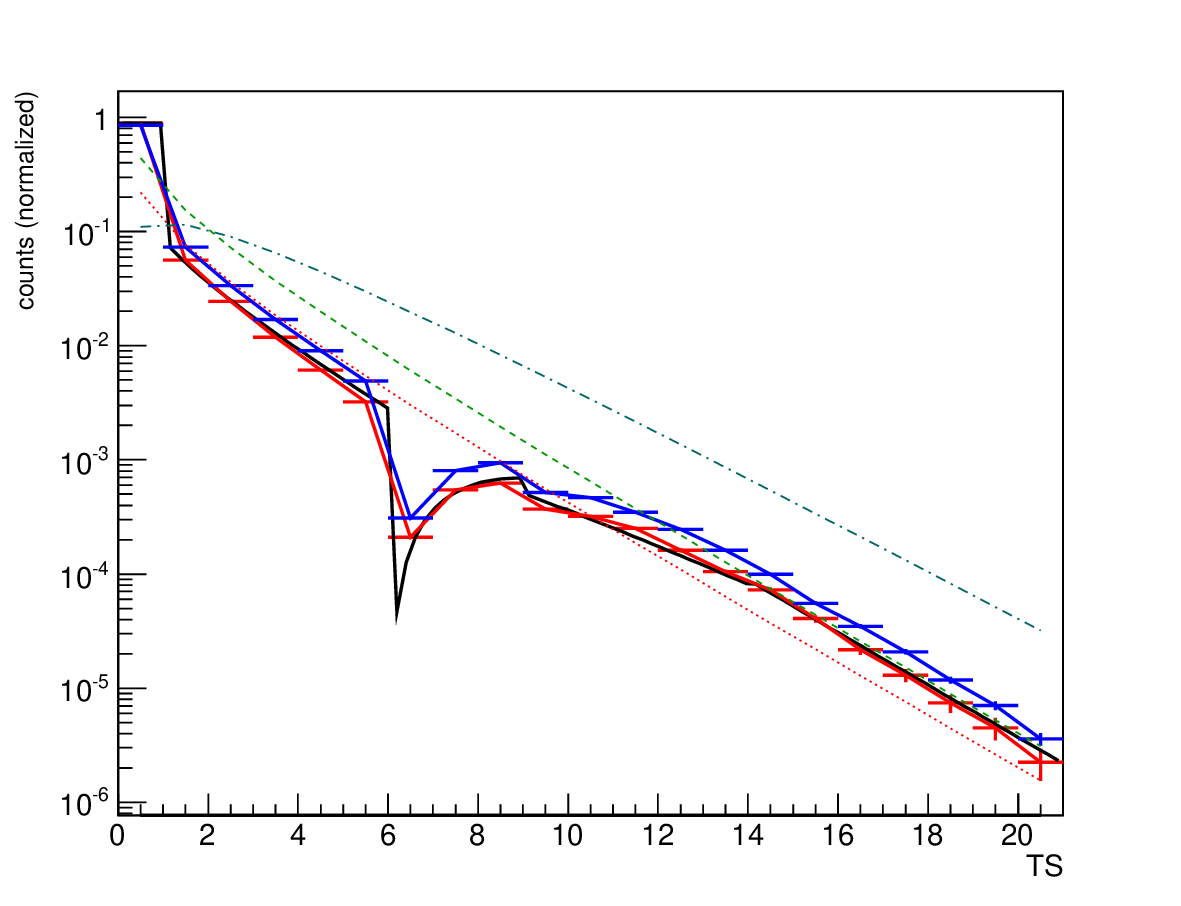}
\centering \caption { {\itshape The effect of steady sources in
the ensemble of models. The blue line is the PDF for Cygnus field
when the null hypothesis for Cygnus X-3 is true, and the red line
is the PDF for an empty Galactic field when the null hypothesis is
true, with the following parameters:  $T_{ss} =4$, $t_{lcl}=5.99147$,
flux and position of source in the alternative hypothesis left
free, $g_{gal}$ and $g_{iso}$ parameters fixed, ICL criterion. Black line is the best fit function described in
Equation \ref{eqn-ICL}. The red dotted
line is the $\frac{1}{2} \chi^2_1$ theoretical distribution, the
green dashed line is the $\chi^2_1$  theoretical distribution, the
Cyan dotted-dashed line is the $\frac{1}{2} \chi^2_3$
distribution.} } \label{fig_OB2400vsCYGX3}
\end{figure}

The following is the analytical expression that can be used with
$t_{ICL} = 9$  where  $T_3=14$:

\begin{equation}
$$
\mbox{$\kappa'''(T_s)$} = \left\{ \begin{array}{ll}
\mbox{ $\delta$ } &\mbox{ if $T_s<1$} \\
\mbox{ $\eta_1 \chi^2_{N_1}(T_s)$ } &\mbox{ if $T_s\ge 1$ and $T_s\le t_{lcl}$ } \\
\mbox{ $\eta_2 \chi^2_{N_2} (T_s - t_{lcl})$ } &\mbox{  if $T_s\ge t_{lcl}$ and $T_s\le T_{ICL}$  } \\
\mbox{ $\eta_3 \chi^2_{N_3} (T_s)$ } &\mbox{  if $T_s\ge T_{ICL}$ and $T_s\le T_3$} \\
\mbox{ $\eta_4 \chi^2_{N_4} (T_s)$ } &\mbox{  if $T_s\ge T_3$ } \\
\end{array} \right.
$$
\label{eqn-ICL}
\end{equation}

\noindent for the following values for the parameters: $N_1=1$,
$N_2=5$, $N_3=5$,  $N_4=1$, $\delta = 0.89 \pm 4.4\times 10^{-4}$,
$\eta_1 = 0.34 \pm 5.1\times 10^{-4}$, $\eta_2 = 4.5 \times
10^{-3} \pm 5.7 \times 10^{-5}$, $\eta_3 = 1.27 \times 10^{-2} \pm
1.7 \times 10^{-4}$, $\eta_4 = 0.91   \pm 3.3 \times 10^{-2}$.
This expression approximates the expected behavior of the
analysis, which shifts gradually from a source location algorithm
with many free parameters near the threshold where the location
contour is large, to an analysis more similar to a fixed-position
analysis at high $T_s$ where the location contour is small.
\abl{Figure \ref{fig_OB2400vsCYGX3} reports the $T_s$ distribution when the
null hypothesis is true produced by maximum likelihood analysis
with the following parameters: $T_{ss} =4$, $t_{lcl}=5.99147$, flux
and position left
free and ICL criterion; red histogram is of a region centered at (l,b)=(160,0)$\degmark$, blue histogram is related to Cygnus region. The black line is the best fit as reported in Eqn. \ref{eqn-ICL}.}
The reported distributions correspond to the standard quick-look
analysis of AGILE data for \abf{empty} Galactic regions. We {\mt
notice } that the \abf{ changes in the selection criterion} {\mt
modify}
the expected $p$-values with respect to the $\frac{1}{2}
\chi^2_3$ theoretical expected distribution (cyan dashed line).

\begin{figure}[!htb]
\centering
\includegraphics[width=9 cm]{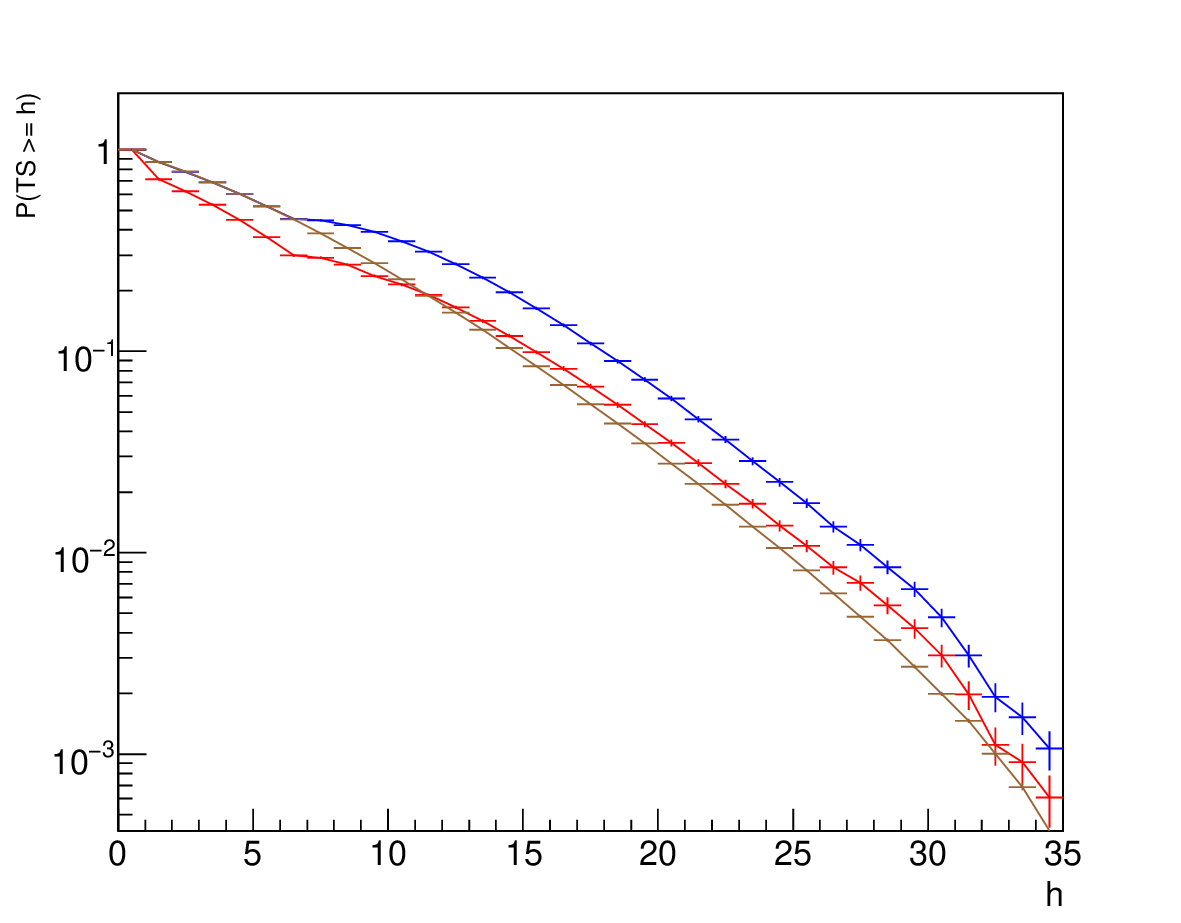}
\centering \caption { {\itshape   The effect of the hypothesis
formulation (trial selection). The  histograms show the $p$-value
distributions in the presence of a simulated source at the
location of (l,b)=(160,0)$\degmark$ with flux $180 \times 10^{-8}$
\phcmsec. Red  histograms have
the following parameters: $T_{ss} =4$, $t_{lcl}=5.99147$, flux and
position of the source left free with ICL rejection criterion. Blue
histograms have $t_{lcl}=5.99147$, flux and position of the source
left free without ICL rejection criterion. Brown histograms have
the flux left free and the position of the source kept fixed. } }
\label{fig_OB2400_4}
\end{figure}

\abf{
The analysis with the position of the source kept fixed yields lower
$p$-values than the analysis with the position allowed to vary,
either with the ICL criterion (red histogram of Figure \ref{fig_OB2400_1BB})
or without (blue histogram). However, the fraction of detections above a
given $T_S$ threshold in the presence of a real source is also lower
when the source position is kept fixed, as shown by the histograms in
Figure \ref{fig_OB2400_4}.  Table \ref{table_det} reports the number of detections
(in $\%$) for some $T_s$ thresholds.
}

\begin {table*}[!htb]
\caption {\em{ $\%$ of detections for some $T_s$ thresholds
(with related $p$-values when the null hypothesis is true)
when the null hypothesis is false with a source at the (l,b)=(160,0)$\degmark$ position
with a simulated flux of $180 \times 10^{-8}$ \phcmsec.   }}
\label{table_det}
\renewcommand{\arraystretch}{1.2}
\begin{tabular}{@{}lllllll}
\hline
& Fixed position & &  Free position $95\%$ & & Free position $95\%$ + ICL  &  \\
$T_s \ge$ & $\%$ detection & $p$-value &   $\%$ detection & $p$-value &  $\%$ detection  & $p$-value \\

\hline
12 & 18.1 & $2.48 \times 10^{-4}$ & 31.3 & $1.21 \times 10^{-3}$ &  22.3 & $4.44 \times 10^{-4}$   \\
16 & 6.2   & $2.95 \times 10^{-5}$ & 16.1 & $2.98 \times 10^{-4}$ &  10.4 & $5.81 \times 10^{-5}$  \\
25 & 0.7   & $2.68 \times 10^{-7}$ & 2.2   & $7.61 \times 10^{-6}$ &    1.3 & $5.26 \times 10^{-7}$   \\
\hline
\end{tabular}
\vskip .1in
Fixed position:  source position fixed, flux free, $T_{ss}=4$, $g_{gal}$ and $g_{iso}$ parameters fixed.\\
Free position 95$\%$: source position free, flux free, $t_{lcl}=5.99147$, $T_{ss}=4$, $g_{gal}$ and $g_{iso}$ parameters fixed.\\
Free position 95$\%$ + ICL: source position free, flux free, $t_{lcl}=5.99147$, $T_{ss}=4$, $g_{gal}$ and $g_{iso}$ parameters fixed, ICL rejection.\\
\end{table*}

\subsection{A complex Galactic region: the case of the Cygnus field }
\label{sec:galcyg}



\begin{figure}[!htb]
\centering
\includegraphics[width=9 cm]{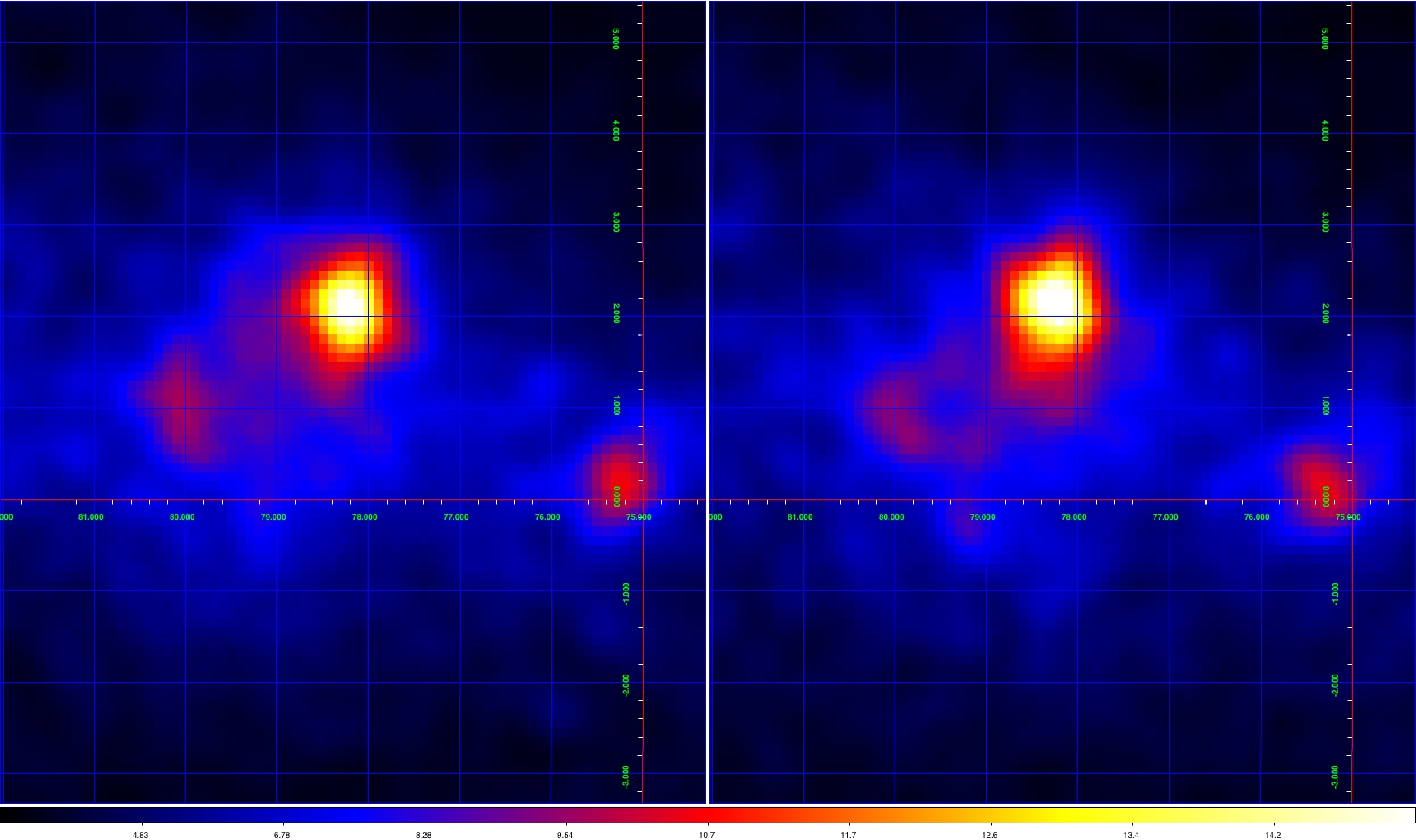}
\centering \caption { {\itshape The binned counts map of the real
(left side) and simulated (right side) Cygnus field {\mt for } an
integration time of 0.5 years (AGILE counts, $E > 100$ MeV). The
real data is taken from July 2007 to October 2009; the map is
centered on (l,b)=(78.75, 0)$\degmark$ in Galactic coordinates
with a bin size of 0.1$\degmark$. } } \label{fig_CF}
\end{figure}

We simulated observations of the Cygnus region both including and without a source
at the position of Cygnus X-3 to test the analysis procedure in a complex case with
nearby point sources and Galactic diffuse emission. We have chosen this field because
this is one of the most complex cases that our analysis procedure must address.
Cygnus X-3 is a well-known microquasar \citep{Giacconi_1967},
showing variable emission at all wavelengths, including repeated $\gamma$-ray
flaring activity above 100 MeV as detected by AGILE \citep{Tavani_2009b}.
This case has been chosen because it shows a great variability in the
$\gamma$-ray energy range and a high correlation with other wavelengths.
The list of simulated sources of the Cygnus region is reported in Table \ref{table_2a}.
In Figure \ref{fig_CF} we show a 0.5 year integration of AGILE data from the Cygnus
region and a simulation of a comparable integration using the same parameters used to
simulate the short trials, demonstrating that the underlying model is sound.

The null hypothesis is that no $\gamma$-ray source coincident with Cygnus X-3 is
present in the AGILE data, while the alternative hypothesis is that a source
coincident with Cygnus X-3 is emitting $\gamma$-rays. The parameters of the
other sources and the diffuse emission coefficients were all kept fixed.

\begin {table*}[!htb]
\caption {\em{List of Cygnus region sources for E $>$ 100 MeV. }}
\label{table_2a}
\renewcommand{\arraystretch}{1.2}
\begin{tabular}{@{}llllllllll} \hline
&  &   & E$>$100 MeV &  &   \\
\textbf{ AGILE Name } & \textbf{ l } & \textbf{ b }  & \textbf{ $\sqrt{TS}$ } & Flux  &  \textbf{Counterpart name} \\
AGL 2021+4029 & 78.24 & 2.16  & 42.1 & 141  $\pm$ 4   & Gamma Cygni \\
AGL 2021+3652 & 75.24 & 0.14  & 23.3 & 67  $\pm$ 3   & PSR J2021+3651 \\
AGL 2030+4129 & 80.11 & 1.25  & 8.1 & 18  $\pm$ 3      & LAT PSR J2032+4127 \\
AGL 2026+3346 & 73.28 & -2.49  & 6.8 & 10 $\pm$ 1.7 & - \\
AGL 2046+5032 & 88.99 & 4.54  & 6.5 & 10 $\pm$ 1.7    & - \\
AGL 2016+3644 & 74.59 & 0.83  & 6.3 & 14 $\pm$ 2.3    & - \\
AGL 2029+4403 & 81.97 & 3.04  & 5.4 & 14 $\pm$ 3      & -\\
AGL 2038+4313 & 82.32 & 1.18  & 5.1 & 15 $\pm$ 3      & - \\
AGL 2024+4027 & 78.56 & 1.63  & 5.0 & 24 $\pm$ 5     & - \\
AGL 2019+3816 & 76.24 & 1.14  & 4.2& 11 $\pm$ 2.4    & - \\
AGL 2036+3954 & 79.47 & -0.56  & 3.4 & 5.0 $\pm$ 1.5 & - \\
\hline
\end{tabular}
\vskip .1in
 The table provides: (1) AGILE name of the sources; (2) (3) the galactic coordinates $l$ and $b$; (4) the statistical significance $\sqrt{T_s}$ of the source detection according to the maximum 
 likelihood ratio test for E$>$100 MeV; (5) the
 period-averaged flux $F$ (\enmev) in $10^{-8}$ \phcmsec for E$>$100 MeV;
 (6) a possible counterpart. We added more  sources in the $\gamma$-ray model compared with the First AGILE Catalog \citep{Pittori_2009}  to take into account a new background event filter (FM3.119).
\end{table*}

If we fix the position of the source as already described in the previous section,
using Equation \ref{eqn_fit1} we find the best fit with
 $\delta = 0.65  \pm 3.6\times 10^{-4}$, $\eta  = 4.6 \times 10^{-1} \pm 4.2 \times 10^{-4}$,
$N_1 = 1$. The two $\eta$ parameters of the cases of empty and complex Galactic fields are
very similar.
Therefore, in the \abh{Tables} \ref{table_final2} and \ref{table_final} we report a single value for the fixed
position analysis which is valid for both cases.

Keeping the position of Cygnus X-3 free and fitting with Equation \ref{eqn-posfree}
we find that $N_1=1$, $N_2=5$, $\delta =0.85 \pm 4.6\times 10^{-4}$, $\eta_1 = 0.46
\pm 6.2\times 10^{-5}$ and $\eta_2 = 6.15\times 10^{-3} \pm 0.4\times 10^{-5}$.
This fitting is appropriate for blind searches for unknown sources in complex Galactic regions.

Keeping the position of Cygnus X-3 free with ICL criterion and fitting with
Equation \ref{eqn-ICL} we find that  $N_1=1$, $N_2=5$, $N_3=5$,  $N_4=1$,
$\delta = 0.84 \pm 2\times 10^{-4}$, $\eta_1 = 0.49 \pm 2.8\times 10^{-4}$,
$\eta_2 = 6.7 \times 10^{-3} \pm 3.2 \times 10^{-5}$,
$\eta_3 = 1.8 \times 10^{-2} \pm 1.0 \times 10^{-5}$,
$\eta_4 = 1.32   \pm 1.9 \times 10^{-2}$. This expression approximates
the expected behavior of the analysis, as already established in the case
of an empty Galactic field (see Section \ref{sec:montecarlo3}).
The reported distributions correspond to the standard quick-look analysis of
AGILE data for complex Galactic regions.

Figure \ref{fig_OB2400vsCYGX3} compares the probability density function for the
Cygnus field (see Table \ref{table_2a}) with this empty Galactic field. As expected,
the effect of {\abe a} more complex
\abf{region} is an increase in the number of false detections.

With the performed \abh{Monte Carlo} simulation we have determined the
$p$-value function for the most common hypothesis formulations.
Based on these simulation we are able to establish the $T_s$ level
for each $p$-value and constrain the false occurrence rate. Table
\ref{table_final2} reports the correspondence between $p$-value and
$T_s$ value for different methods of analysis, in addition to the
theoretical reference for $\chi^2_1$, $1/2 \chi^2_1$ and $  1/2
\chi^2_3$.



\begin {table*}[!htb]
\caption {\em{ Correspondence between $p$-value and $T_s$ value for different methods of analysis.  The first column reports the $p$-value, the following columns report the corresponding $T_s$value. In particular, the fifth  column reports the corresponding $T_s$ value for Galactic regions with the position of the source of the alternative hypothesis kept fixed, the sixth and 	last   columns report the corresponding $T_s$ value for Galactic regions (first number for empty regions, second number for complex regions) with the position of the source of the alternative hypothesis kept free.}}
\label{table_final2}
\renewcommand{\arraystretch}{1.2}
\begin{tabular}{@{}lllllll}
\hline
  &   &  & & $T_s$  &   &    \\
$p$-value & $\chi^2_1$ & $1/2 \chi^2_1$ & $  1/2 \chi^2_3$ & Fixed position  & Free position 95$\%$ & Free position 95$\%$ + ICL   \\
\hline
$10^{-2}$ & 6.63    &  5.41   &   9.84 & 5.29   &  4.78-5.27 & 4.78-5.41 \\
$10^{-3}$ & 10.83  &  9.55   & 14.80 & 9.42   &  12.60-13.90 & 9.89-10.84 \\
$10^{-4}$ & 15.14  &  13.83 & 19.66 & 13.70 &  18.81-19.91 & 14.97-15.67 \\
$10^{-5}$ & 19.52  &  18.19 & 24.47 & 18.06 &  24.36-25.40 & 19.35-20.05 \\
$10^{-6}$ & 23.93  &  22.60 & 29.23 & 22.46 &  29.66-30.66 & 23.76-24.47 \\
$10^{-7}$ & 28.37  &  27.03 & 33.98 & 26.90 &  34.81-35.79 & 28.21-28.92 \\
$10^{-8}$ & 32.83  &  31.49 & 38.71 & 31.36 &  39.87-40.84 & 32.67-33.39 \\
$10^{-9}$ & 37.32 & 35.97 & 43.42 & 35.84   &  44.87-45.82 & 37.16-37.87 \\
\hline
\end{tabular}
\vskip .1in
 Fixed position:  source position fixed, flux free, $T_{ss}=4$, $g_{gal}$ and $g_{iso}$ parameters fixed.\\
Free position 95$\%$: source position free, flux free, $t_{lcl}=5.99147$, $T_{ss}=4$, $g_{gal}$ and $g_{iso}$ parameters fixed.\\
Free position 95$\%$ + ICL: source position free, flux free, $t_{lcl}=5.99147$, $T_{ss}=4$, $g_{gal}$ and $g_{iso}$ parameters fixed, ICL rejection.\\
\end{table*}

\subsection{Deviation from  the nominal distribution}

\abf{
In the following we report the effect that uncertainties in the analysis parameters produce on
the shape of the distributions. The analyses were performed for the case of the complex Galactic region.}

\subsubsection{The effect of the $t_{lcl}$ parameter}
\begin{figure}[!htb]
\centering
\includegraphics[width=9 cm]{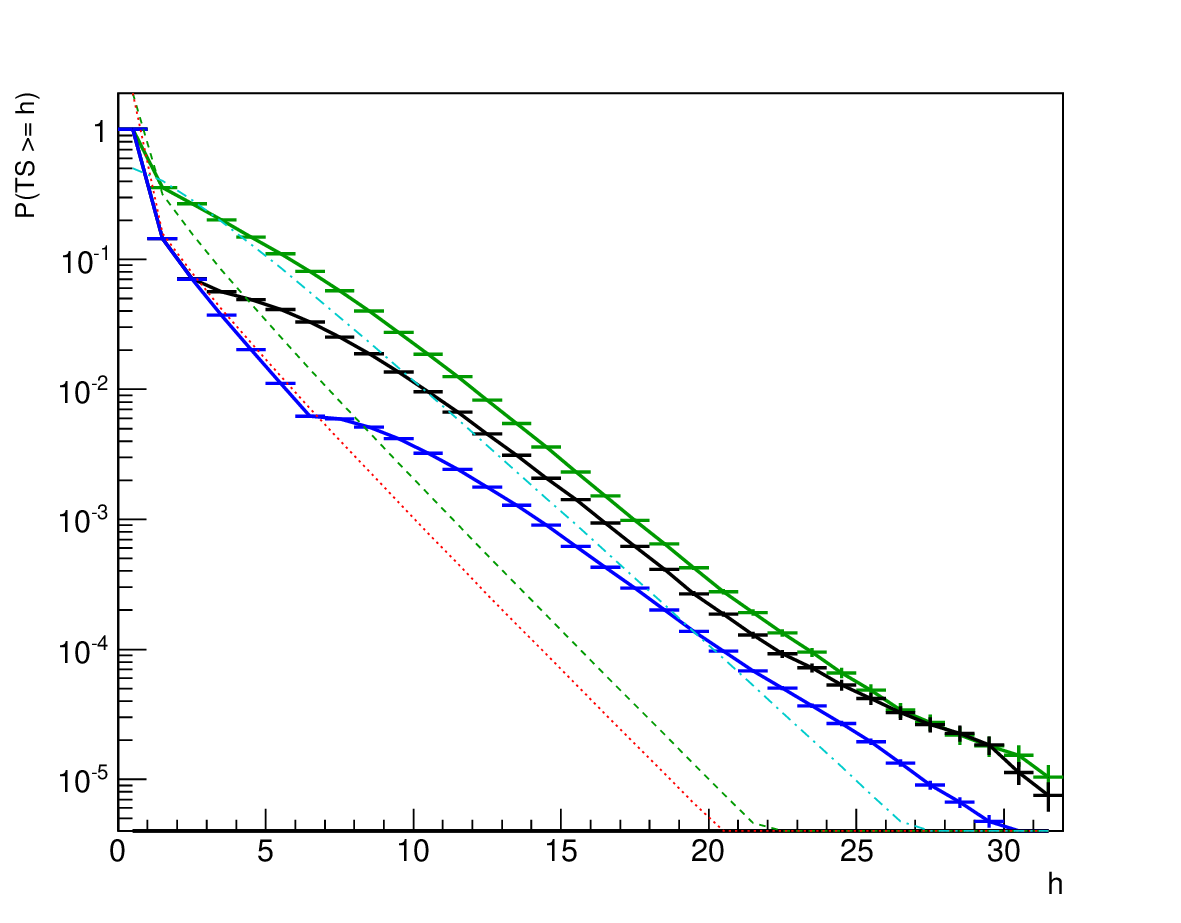}
\centering \caption { {\itshape The $p$-value distributions change
as a function of the $t_{lcl}$ parameter. Simulations of the
Cygnus region with no source present at  Cygnus X-3 position
were analyzed with different $t_{lcl}$ values without applying the
ICL rejection criterion. The flux and position of the hypothetical
source at the Cygnus X-3 were allowed to vary, the $g_{gal}$ and
$g_{iso}$ parameters were kept fixed, and $T_{ss}=4$. Green
histogram: $t_{lcl}=0$; black histogram: $t_{lcl}=2.29575$ (corresponds to a $68\%$ confidence level for two degrees of freedom); blue
histogram: $t_{lcl}=5.99147$. The red dotted line is the $\frac{1}{2}
\chi^2_1$ theoretical distribution, the green dashed line is the
$\chi^2_1$  theoretical distribution, and the cyan dotted-dashed
line is  $\frac{1}{2} \chi^2_3$ distribution.} }
\label{fig_CYGX3_2A1}
\end{figure}

We performed additional maximum likelihood analyses using
different values of the $t_{lcl}$ parameter, including the case of
$t_{lcl}=0$. In the case of a blind search for unknown sources,
the maximum likelihood estimator can be used as a source finder,
instead of an hypothesis validator, by setting $t_{lcl}=0$. In
this case the optimization of the position is performed regardless
of the $T_s$ found in the first step with fixed position. The
resulting $p$-value distributions  are shown in Figure
\ref{fig_CYGX3_2A1} (with $t_{lcl}=0$ shown in green) including
all trials without ICL rejection. As expected, higher $t_{lcl}$
values correspond to lower $p$-values because the optimization of
the position starts for higher $T_s$ values in the first loop of
the maximum likelihood procedure.

\subsubsection{The effect of the radius of analysis}
\begin{figure}[!htb]
\centering
\includegraphics[width=9 cm]{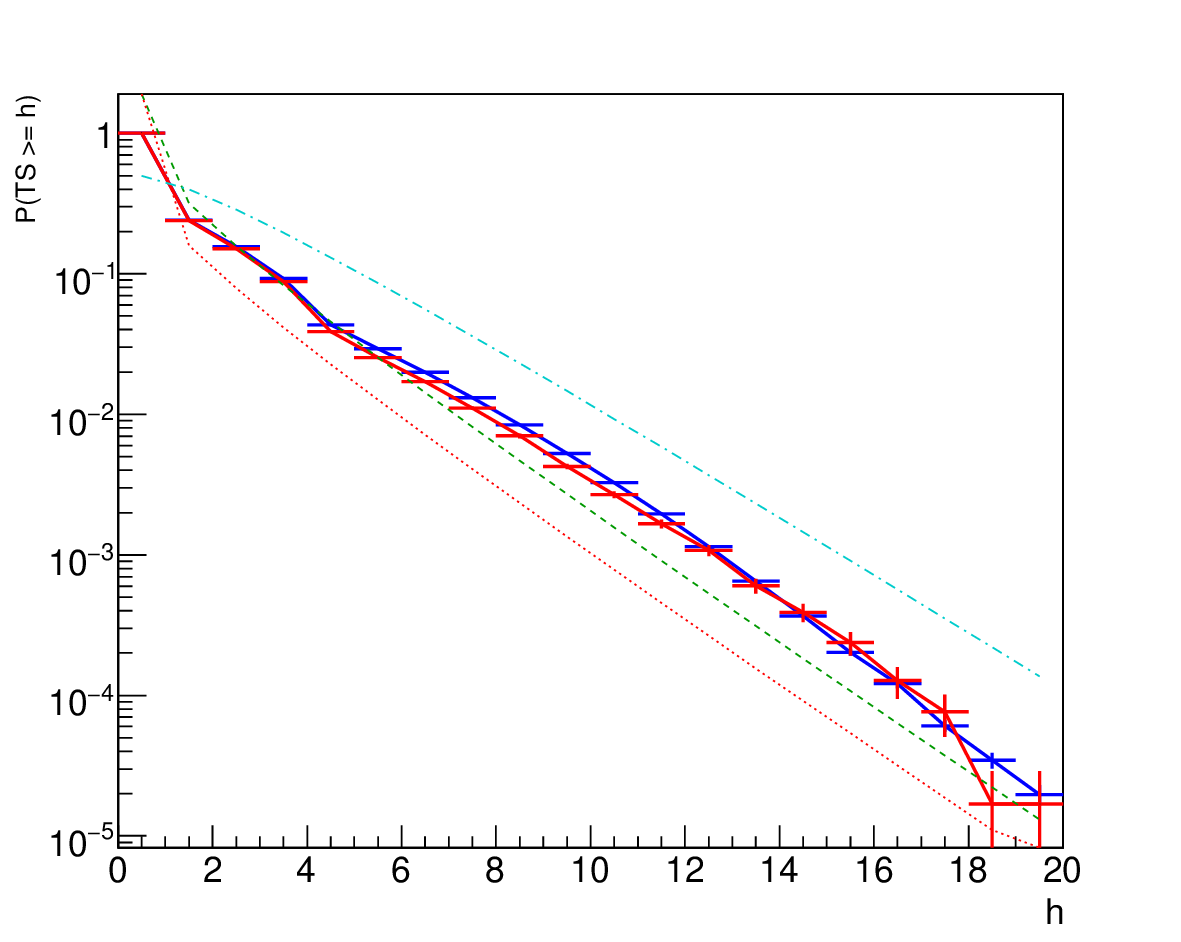}
\centering \caption { {\itshape Comparison between different radii
of analysis. The histograms are the $p$-value distributions for
the Cygnus field when the null hypothesis for Cygnus X-3 is true
with the following parameters: $T_{ss} =4$, $t_{lcl}=0$, flux and
position of Cygnus X-3 left free, ICL rejection applied. Red
histogram:radius of analysis $=10\degmark$; blue histogram: radius
of analysis $=5\degmark$. The red dotted line is the $\frac{1}{2}
\chi^2_1$ theoretical distribution, the green dashed line is the
$\chi^2_1$ theoretical distribution, and the cyan dotted-dashed
line is the $\frac{1}{2} \chi^2_3$ distribution. } }
\label{fig_CYGX3_12}
\end{figure}

Figure \ref{fig_CYGX3_12} shows that no appreciable differences are produced
by changing the radius of analysis. The comparison was performed for the case
of $t_{lcl}=0$ applying the ICL criterion, but should also be valid for
analyses using the other \abh{criteria}.

\subsubsection{The effect of of keeping  $g_{gal}$ and $g_{iso}$ free or fixed}
\begin{figure}[!htb]
\centering
\includegraphics[width=9 cm]{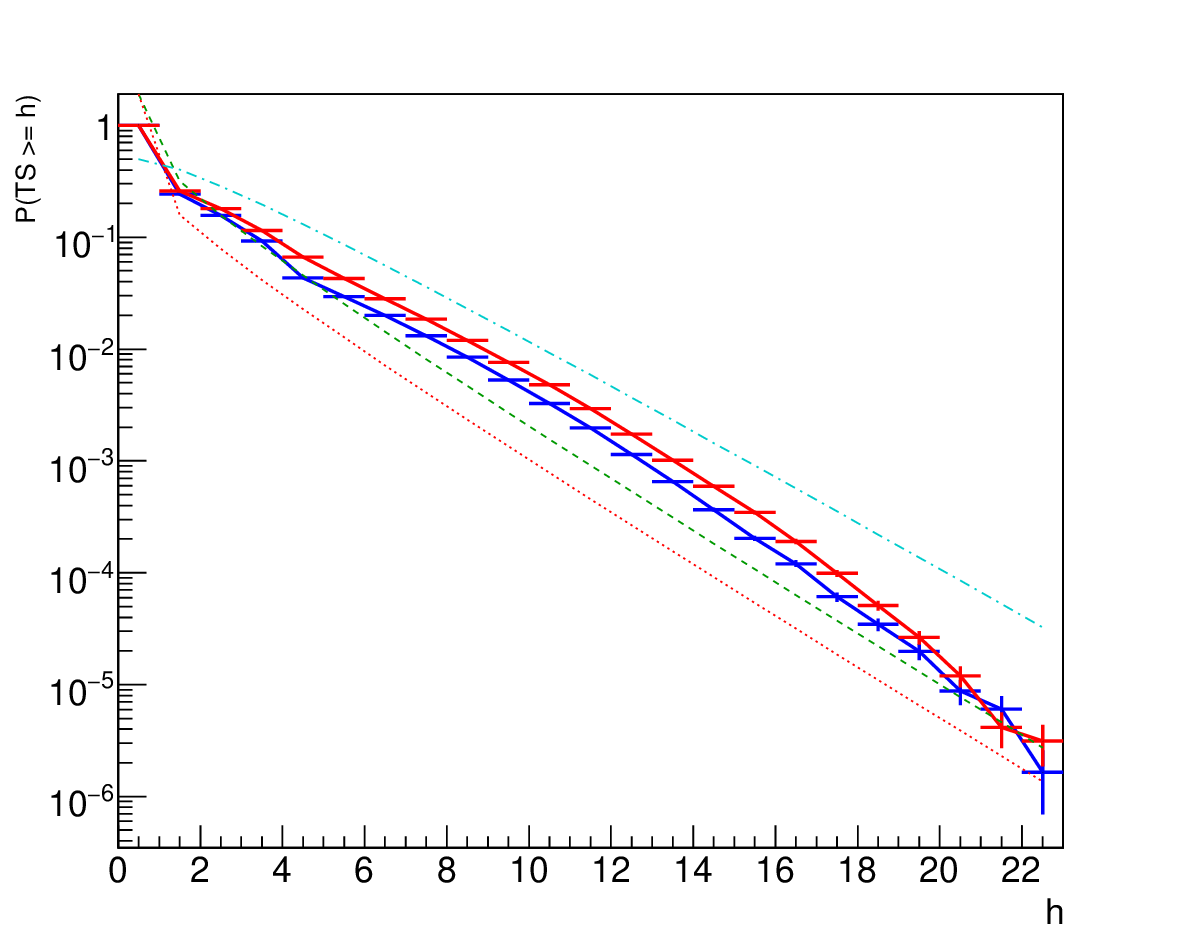}
\centering \caption { {\itshape Comparison between $g_{gal}$ and $g_{iso}$ parameters free or fixed. \abl{The blue} histogram is the $p$-distribution for \abl{the} Cygnus field when the null hypothesis for Cygnus X-3 is true with the following parameters: $T_{ss} =4$, $t_{lcl}=0$, flux and position of Cygnus X-3 left free, $g_{gal}$ and $g_{iso}$ parameters fixed. \abl{The red} histogram has the same parameters but with $g_{gal}$ and $g_{iso}$ parameters left free. The red dotted line is the $\frac{1}{2} \chi^2_1$ theoretical distribution, the green dashed line is the $\chi^2_1$  theoretical distribution, the cyan dotted-dashed line is the $\frac{1}{2} \chi^2_3$ distribution. } }
\label{fig_CYGX3_8}
\end{figure}

In Figure \ref{fig_CYGX3_8} we show the effect of keeping 
$g_{gal}$ and $g_{iso}$ parameters fixed (blue) and free (red): the $p$-values
\abl{with free parameters are larger}. This result is expected
because fixing these parameters reduces the range of possible hypotheses explored.
The comparison was performed for the case of $t_{lcl}=0$ applying the ICL criterion, but should also be valid for analyses using the other \abh{criteria}.

\subsubsection{The effect of unmodeled point sources}

\begin{figure}[!htb]
\centering
\includegraphics[width=9 cm]{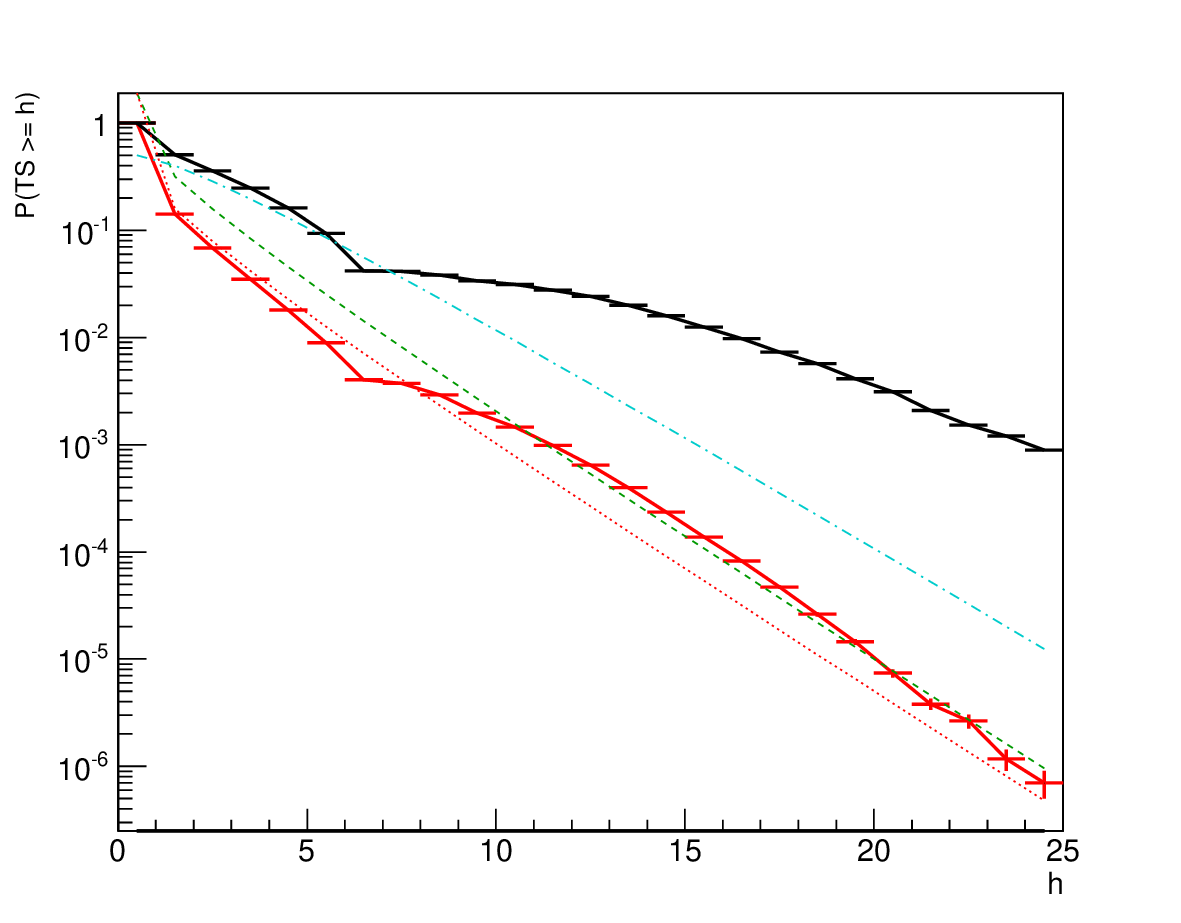}
\centering \caption { {\itshape The effect of unmodeled sources.
The histograms are the $p$-value distribution for the Cygnus field
when the null hypothesis for Cygnus X-3 is true with the following
parameters: $T_{ss} =4$, $t_{lcl}=5.99147$, flux and position of
Cygnus X-3 left free, $g_{gal}$ and $g_{iso}$ parameters fixed.
The red histogram shows the result when all the simulated sources
are included in the models, while the black histogram shows the
result when only the source at Cygnus X-3 position is included.
The red dotted line is the $\frac{1}{2} \chi^2_1$ theoretical
distribution, the green dashed line is the $\chi^2_1$ theoretical
distribution, the cyan dotted-dashed line is the $\frac{1}{2}
\chi^2_3$ distribution.} } \label{fig_CYGX3_90}
\end{figure}

We performed a Monte Carlo simulation in which the simulated data
contain all of the sources listed in Table \ref{table_2a},
followed by a maximum likelihood analysis with models containing
only a point source at the location of Cygnus X-3, in order to
evaluate the effect of nearby unmodeled point sources. The
resulting $p$-value distribution is shown in Figure
\ref{fig_CYGX3_90}, where the red histogram is the $p$-value
distribution with the analysis models containing all the sources
used in the simulation and the black histogram is the
$p$-distribution with only the source at the Cygnus X-3 location.
\abl{This illustrates that it is critical to model existing sources correctly}.

\subsubsection{The effect of errors in the diffuse emission: estimating using $g_{gal}$}

\begin{figure}[!htb]
\centering
\includegraphics[width=9 cm]{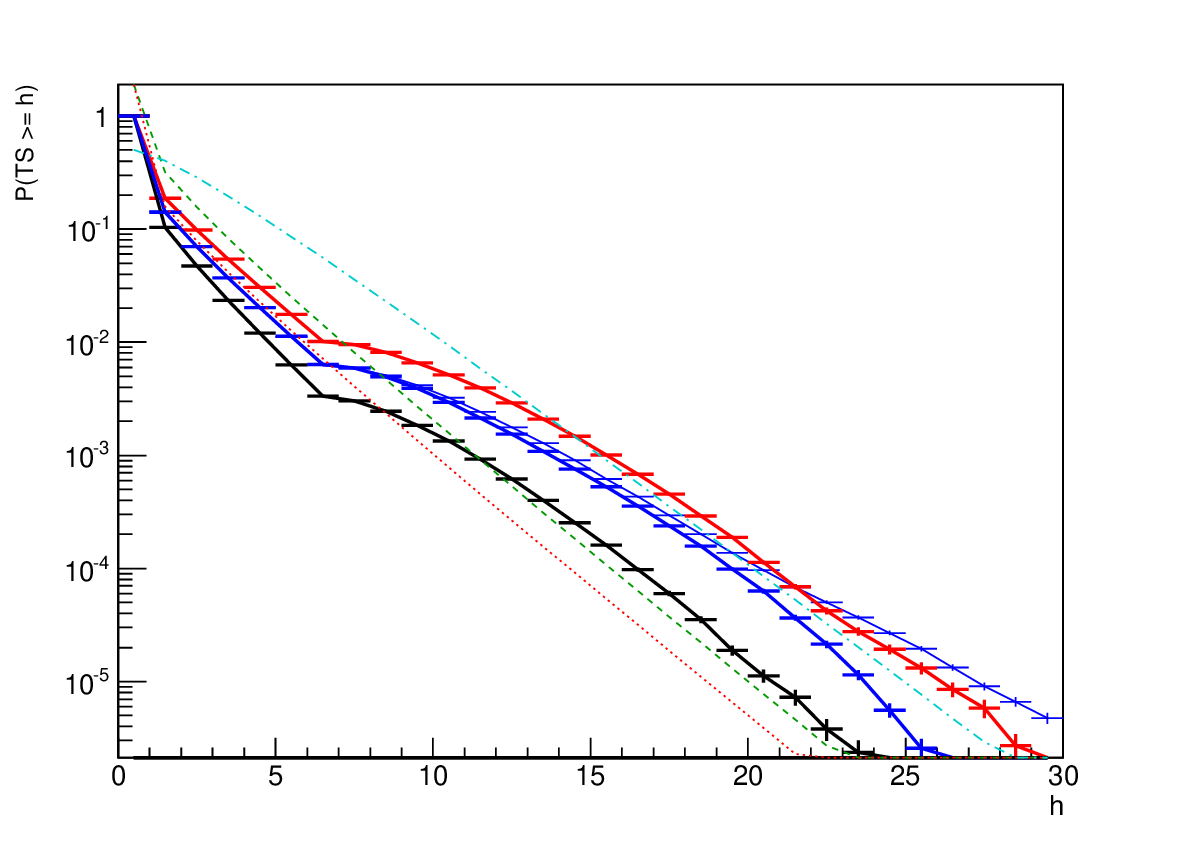}
\centering \caption { {\itshape \abl{The effect of a poor estimation of
the Galactic diffuse emission. The histograms are the $p$- value
distributions for the Cygnus field when the null hypothesis for
Cygnus X-3 is true with the following parameters: $T_{ss} = 4$, $t_{lcl} = 5.99147$, flux
and position of Cygnus X-3 left free.  The true value of $g_{iso}$ in the
simulated data is $7.7$, the value of $g_{gal}$ used in the analysis is fixed to $0.67$. Blue thin line: simulated $g_{gal}=0.67$, $g_{iso}$  fixed to $7.7$; blue thick 
line: simulated $g_{gal}=0.67$, $g_{iso}$ \abh{left} free; red thick 
line: simulated $g_{gal}=0.67*1.1$, $g_{iso}$ \abh{left}  free; black thick line: simulated $g_{gal}=0.67*0.89$, $g_{iso}$ \abh{left}  free. The red
dotted line is the $\frac{1}{2} \chi^2_1$ theoretical
distribution, the green dashed line is the $\chi^2_1$  theoretical
distribution, the cyan dotted-dashed line is the $\frac{1}{2}
\chi^2_3$ distribution.}} } \label{fig_CYGX3_900}
\end{figure}

\abl{We performed a preliminary investigation of the effect of systematic errors
in the diffuse emission model on the analysis results by fixing the $g_{gal}$
parameter during the analysis at a value $g_{gal} = 0.67$ different from the one used in
simulating the data;   $g_{iso}= 7.7$ is simulated \abh{but the $g_{iso}$ parameter is left free to adapt} during the analysis.  Figure \ref{fig_CYGX3_900} reports the  $p$- value distributions for the Cygnus field when the null hypothesis for
Cygnus X-3 is true where the results of three analyses are compared;
one with simulated $g_{gal}=0.67$ (blue thick lines), one with simulated  $g_{gal}=0.67*1.1$
(red thick lines) and one with simulated  $g_{gal}=0.67*0.89$ (black thick lines). Table \ref{table_iso} reports the resulting calculated $g_{iso}$.}

\begin {table}[!htb]
\caption {\em{ The mean value of the calculated $g_{iso}$ parameter }}
\label{table_iso}
\renewcommand{\arraystretch}{1.2} 
\begin{tabular}{|cc|cc|}
\hline  
Simulations &    & Analysis   & \\
\hline 
 $g_{gal}$ &  $g_{iso}$    & $g_{gal}$ fixed &  $g_{iso}$ calculated   \\
\hline 
$0.67*0.89$ & 7.7 & 0.67 & $6.55  \pm  1.46$     \\
$0.67$ & 7.7 & 0.67 &  $7.55 \pm  1.52$     \\
$0.67*1.1$  & 7.7 & 0.67 & $8.52 \pm  1.59$     \\
\hline
\end{tabular}
\end{table}

As expected, $g_{iso}$ \abh{moves} up if $g_{gal}$ is too small and vice versa. If $g_{gal}$ is under-estimated, then
 the number of false detections when the source is absent \abh{increases} (see the red lines
of Figure \ref{fig_CYGX3_900}), because background photons are assigned
to the source. If
the $g_{gal}$ parameter is over-estimated, then  the number of false
detections when the source is absent decrease (see black line of Figure \ref{fig_CYGX3_900}) because \abh{the diffuse model is already too high at the position of Cygnus X-3}.

\section{Pre-trials and post-trials significance}
\label{sec:pretrials} We {\mt have seen how the \abh{Monte Carlo}
simulations can be used } to characterize the $T_s$ distributions
produced by the AGILE-GRID maximum likelihood analysis procedure.
{\mt In the end, we find} the probability, or $p$-value, of
finding a false positive detection (rejecting the null hypothesis
when it is true) in a single observation.

In practice, for each region of the sky we perform many trials during the daily monitoring
to search for transient $\gamma$-ray events. The probability of obtaining a single false
detection over a large number of trials is therefore much higher than $p$. \abg{In the AGILE context we perform two kinds of analysis for each analyzed map: 
\begin{enumerate}
\item blind search for unknown sources: we search for more than one source at a time with free positions;
\item searches from a list of known sources: we search for more than one source at a time with fixed source positions.
\end{enumerate}
}
\abg{Let $K = M \cdot N$ the number of independent
trials, where N is the number of maps and M is the number of unknown sources in the first case or the number of sources in the predefined list in the second case. If we have only one source in both cases $M=1$.}

Since the probability of \textit{not} making a false positive error in a single trial
is $1-p$, the probability of not making any false positive error is $(1-p)^K$ (type-I error), so the probability of making at least one false
positive error is $\pi = 1-(1-p)^K$. This is defined as the post-trial probability,
also referred to as the \textit{experiment-wide error rate}, while $p$ is denoted as
the pre-trial probability, or \textit{comparison-wise error rate}.
For an experiment-wide false positive rate of $\pi$, we can constrain the comparison-wise
error rate with \abf{the Dune-\^{S}id\'ak correction $p \leq 1 - (1-\pi)^{1/K}$.}

Let us consider   \abg{a typical AGILE} case in the context of Galactic $\gamma$-ray transients with
the position of the source left free \abg{and with a single source with flux and position free for each map $(M=1)$: usually we keep fixed the position and flux of known sources assuming that only one source is in a flaring state in our map (we can reduce the size of the map to accomplish this)}. If we search for a single flaring source once every
two days ($N=182$ for 1 year of observations in AGILE spinning mode), if we can accept one
false detection \abf{during the year} $\pi \leq 1/N$, this implies threshold of $T_S \simeq 17.9$.
If we can accept a false detection once every 2 years, the threshold is $T_S \simeq 20.5$.

\abg{Table \ref{table_final} reports the post-trial significance expressed in Gaussian standard deviations for some values of K.}

\begin {table*}[!htb]
\caption {\em{Post-trial significance expressed in Gaussian standard deviations ($\sigma$).
The first column reports the pre-trial significance, the second column reports the corresponding
$p$-value, the third  column reports the corresponding $T_s$ value for Galactic regions with the
position of the source of the alternative hypothesis kept fixed, the fourth and fifth   columns report
the corresponding $T_s$ value for Galactic regions (first number for empty regions, second
number for complex regions) with the position of the source of the alternative hypothesis
\abh{left free}, the last two columns report the post-trial significance for K=180 and K=360 trials.}}
\label{table_final}
\renewcommand{\arraystretch}{1.2}
\begin{tabular}{@{}llllllll} \hline
$\sigma$ pre-trial & $p$-value & $T_s$ & & & $\sigma$ post-trial \\
& & Fixed position & Free position 95$\%$ & Free position 95$\%$ + ICL & K=180 & K=360 \\
3 & $1.35 \times 10^{-3}$ & 8.88 & 11.66-13.03 & 9.08-10.05 & 0.78 & 0.29 \\
4 & $3.17 \times 10^{-5}$ & 15.87 & 21.63-22.69 & 17.14-17.85 & 2.53 & 2.28 \\
5 & $2.86 \times 10^{-7}$ & 24.87 & 32.47-33.46 & 26.17-26.88 & 3.88 & 3.71 \\
6 & $9.21 \times 10^{-10}$ & 36.00 & 45.05-46.00 & 37.31-38.03  & 5.10 & 4.97 \\
\hline
\end{tabular}
\vskip .1in
 Fixed position:  source position fixed, flux free, $T_{ss}=4$, $g_{gal}$ and $g_{iso}$
 parameters fixed.\\
Free position 95$\%$: source position free, flux free, $t_{lcl}=5.99147$, $T_{ss}=4$, $g_{gal}$
and $g_{iso}$ parameters fixed.\\
Free position 95$\%$ + ICL: source position free, flux free, $t_{lcl}=5.99147$, $T_{ss}=4$,
$g_{gal}$ and $g_{iso}$ parameters fixed, ICL rejection.\\
\end{table*}








\section{Probability of sky-position coincidence in non-overlapping time intervals}
\label{sec:multiple}

In this section we generalize the analysis performed in the previous sections to calculate the
probability of two or more detections of a flaring source performed in the same sky position in
different independent time intervals. 
Multiple detections of a source can have a low probability
of being consistent with the null hypothesis even when the individual detections are at a low level
 of $T_s$.
In order to assess the statistical significance of our detections,
we consider the post-trial probability of flare occurrence. We
distinguish two cases:

\begin{enumerate}
\item  the case of a single flare episode originating from a
specific source within a given error box (that we define as
``single independent occurrence'' or ``single post trial
occurrence"); \item the case of repeated flaring episodes
originating from a specific source with a given error box  (that
we call here ``repeated post-trial flare occurrence''.
\end{enumerate}

\noindent For each individual {\mt AGILE} detection, we can
calculate the post-trial significance of the \textit{single
independent occurrences},  which does not take into account the
history of repeated occurrences.

We can {\mt then} combine the history of the sky region and
establish the probability of repeated flaring episodes from the
same sky position. We calculate the post-trial significance for
\textit{repeated flare occurrences} at the source error-box
position as follows. 

\abg{If we perform one trial for each map ($M=1$, we use a list with only one source;  this means that
each independent time period is a single trial), the} chance probability
of
having $k$ or more detections \abl{over} $N$ maps at a specific site with a $T_s$
statistic satisfying $T_s \ge h$  is given by $ P(N, X \ge k)
= 1 - \sum_{j=0}^{k-1}
\left(\begin{array}{c}N\\j\end{array}\right) p^j (1-p)^{N-j}$
where $p = p(h)$ is the $p$-value corresponding to the $h$ value  given by Equation \ref{eq_A},
$P(N, X=j) = \left(\begin{array}{c}N\\j\end{array}\right) p^j (1-p)^{N-j}$ is the probability of
exactly $j$ detections in $N$ maps and $P(N, X < k) =  \sum_{j=0}^{k-1}
\left(\begin{array}{c}N\\j\end{array}\right) p^j (1-p)^{N-j}$ is the probability
of fewer than $k$ detections \abe{at a specific position} in $N$ maps.

\abg{If we perform $M$ trials in different positions of $N$ maps, where $M$ is the number of known or unknown sources in a predefined list, the chance of having $k$ or more detections above the level $h$ in any of the sites with a $T_s$
statistic satisfying $T_s \ge h$  is given by  $P_M(N, X \ge k)  = 1 - \left( \sum_{j=0}^{k-1} \left(\begin{array}{c}N\\j\end{array}\right) p^j (1-p)^{N-j} \right)^M$ where $P_M(N, X \ge k) =  1 - P(N, X < k)^M = 1 - (1 - P(N, X \ge k))^M$.}

The choice  of $p(h)$ depends on our assumptions: if the flare comes from \abl{a} known source we use Equation \ref{eqn-ICL}, if the flare comes from \abl{a} previously unknown source we use Equation \ref{eqn-posfree} (the case without ICL criterion). 

Let us consider a simple case in the context of Galactic
$\gamma$-ray transients, using the $p$-value function in the case
of a complex Galactic region  with the position of \abg{a single} source left
free. If we detect one flare per year at a specific position of
the Galactic plane with $T_s > 16$, and we produce 182 maps per
year (once every two days with an integration time of two days),
then after the first year the global post-trial significance is
2.16 $\sigma$, after the second year it is 3.31 $\sigma$, and
after the third year it is 4.16 $\sigma$. Transient sources, as
long as they recur, enable us to have more confidence in their
detection as integration time increases.

It might seem that \abh{this} approach adds a bias to \abh{the}
\textit{global significance} of a detection of a flaring source,
because it may happen that some flares from nearby sources can be
"counted" together due to the extension of the $95\%$ contour
confidence level. In doubtful cases only a more detailed analysis can exclude
 this.

\section{Conclusions}
\label{sec:conclusion}

\abf{ We have performed {\mt extensive} \abh{Monte Carlo} simulations to
characterize the maximum likelihood ratio test for the AGILE-GRID
instrument  in the context of short timescale (1-2 days) flaring
$\gamma$-ray sources, both in  extra-Galactic and Galactic fields.
In the case of  Galactic fields, we have simulated both a simple
(without steady sources)
 and a complex Galactic region. With these simulations we have calibrated both the $T_s$
  distributions (pre-trial significance) and the related false occurrence rate.}

After the introduction of
 the post-trial significance, we calculated the {\mt post-trial probabilities for
 single and multiple occurrences. In particular, we calculated the}
  probability of two or
 more detections of a flaring source at the same sky
 position in different independent time intervals.
 With this approach, we take into account the presence of many flaring episodes {\mt originating}
  from the same sky region, combining its history and adding information not present
 in the single episode and in the post-trial evaluation. We call this
 ``repeated post-trial flare occurrence''.

In this paper we have provided a method for converting the $T_s$
produced by any of the various methods made available by the AGILE
analysis software into a probability. This information {\mt can}
be used by anyone who performs analysis on {\mt GRID data through
the AGILE Guest Observer Program. 

\section{Acknowledgements}
The AGILE Mission is funded by the Italian Space Agency (ASI) with
scientific and programmatic participation by the Italian Institute
of Astrophysics (INAF) and the Italian Institute of Nuclear
Physics (INFN). We acknowledge financial contribution from the agreement ASI-INAF I/009/10/0.

\end{document}